\documentclass[epj]{svjour}
\usepackage{amsmath}
\usepackage{amssymb}
\usepackage[cp1251]{inputenc}
\usepackage[T2A]{fontenc}
\usepackage{latexsym}
\usepackage[dvips]{graphicx}
\usepackage[english]{babel}

\def\bge{\begin{equation}}
\def\ene{\end{equation}}
\def\bgea{\begin{eqnarray}}
\def\enea{\end{eqnarray}}
\def\nn{\nonumber}

\begin{document}
\sloppy
\title{Radiative decays with light scalar mesons and singlet-octet mixing in ChPT}
\author{S.~A.~Ivashyn 
and
A.~Yu.~Korchin 
}
\institute{ Institute for Theoretical Physics, NSC ``Kharkov Institute of
Physics and Technology'', Kharkiv 61108, Ukraine,
\\ \email{ivashyn@kipt.kharkov.ua,~korchin@kipt.kharkov.ua} }
\date{Received: 18 July 2007 / Revised version: 27 November 2007}

\abstract{ We study different types of radiative decays involving
$f_0(980)$ and $a_0(980)$ mesons within a unified ChPT-based
approach at one-loop level. Light scalar resonances which are seen
in $\pi\pi$, $\pi\eta$, $K\bar{K}$ channels of $\phi(1020)$
radiative decays and in $J/\psi$ decays are responsible for key
questions of low-energy dynamics in the strong interaction sector,
and decays $\phi(1020) \to \gamma a_0(980)$, $\phi(1020) \to
\gamma f_0(980)$, $a_0(980) \to \gamma\gamma$, $f_0(980) \to
\gamma \gamma$ are of interest for current experimental programs
in J\"ulich, Frascati and Novosibirsk. From theoretical point of
view it is important to verify whether light scalar mesons are
members of some flavor octet or nonet. We find a value of mixing
angle dictated by consistency with experiment and coupling
structures of ChPT Lagrangian. Decay widths $f_0(980)/
a_0(980)\to\gamma \rho(770)/\omega(782)$, which are not studied
experimentally yet, are predicted. We also obtain several
relations between widths, which hold independently of coupling
constants and represent a fingerprint of the model.
} 
\maketitle
\section{Introduction}
\label{intro}


The scalar mesons $a_0(980)$ ($I^G(J^{PC}) = 1^-(0^{++})$) and
$f_0(980)$ ($I^G(J^{PC}) = 0^+(0^{++})$) have been discussed for
more than 30 years. The shape of $\pi\pi$ (and $\pi\eta$)
invariant mass distribution in different reactions points to these
resonances. The promising source of information on scalar mesons
are radiative decays in which scalar mesons are involved. Much
experimental attention has been paid so far to processes
$\phi(1020) \to \gamma a_0$~\cite{Aloiso02C} and $\phi(1020) \to
\gamma f_0$~\cite{KLOEresults} (see
also~\cite{SNDresults,CMD2results}) due to motivation put forward
in~\cite{Achasov_Ivanchenko}. Recent example of a model describing
such features in the rare $\phi \to \gamma S \to \gamma \pi\eta$
($\gamma\pi\pi$) decays is chiral approach with derivative
couplings ~\cite{Harada06}. Among other well-known processes
involving scalar resonances one can think of $J/\psi \to \phi
f_0(980) \to \phi \pi\pi$ (and $\to \phi K\bar{K}$) studied at
BES~\cite{Ablikim:2004wn} and nucleon-nucleon (as well as
deuteron-deuteron) reactions with various hadronic final states.
The transitions $a_0 \to\gamma\gamma$ and $f_0 \to\gamma\gamma$
are relevant for numerous reactions, where two-photon interactions
produce miscellaneous hadronic final states. Many experiments
involving $\gamma\gamma \to \pi\pi$ (or $ \pi\eta$) have been
carried out or are being planned.

The properties of scalar mesons are not well understood. Nevertheless the
dominant decay channels are known to be $\pi\pi$ for $f_0$ meson and
$\pi\eta$ for $a_0$, and the total widths are in between $40$ and $100$
$\text{MeV}$. The decays to strange mesons $a_0 \to K \bar{K}$ and $f_0
\to K \bar{K}$ are dynamically allowed, though the masses of $a_0(980)$
and $f_0(980)$ may lie slightly below the $K\bar{K}$ threshold. The masses of
$a_0(980)$ and $f_0(980)$ are approximately equal.

The internal structure of light scalar mesons is also not clear.
Recent advances in understanding of their structure are presented
in review~\cite{Bugg:2004xu}. Most of studies show that light
scalar meson structure can not be explained in simple quark
models. This is probably related to a special role played by these
mesons in low-energy dynamics of strong
interaction~\cite{Pennington1,Pennington2}. Namely, scalar fields
can be viewed as the Higgs sector of strong interaction, i.e.
their non-zero vacuum expectation value leads to chiral symmetry
breaking and directly reflect the structure of quark condensate in
Quantum Chromodynamics~(QCD). Some authors emphasize proximity of
the $K\bar{K}$ threshold to $a_0$ and $f_0$ masses that favors
presence of the molecular $K\bar{K}$
component~\cite{Weinstein:1990gu} (for recent calculations
implementing molecular $K\bar{K}$ model
see~\cite{Lemmer:2007qp,Hanhart:KKMol-phi,Hanhart:KKMol-gg}.)

Another approach to light scalar meson features is unitarized
ChPT~\cite{Oller:2000ma}. There the inverse amplitude
method~\cite{Dobado:1996ps} was  employed to describe elastic
$\pi\pi$, $\pi\eta$, $K\eta$ and $K\bar{K}$ scattering data. The
radiative decays in question were evaluated through final state
interaction of scattered particles. The unusual large-$N_c$
behavior of scalar resonances was recently summarized
in~\cite{Jaffe:2007id} (see also references therein and original
paper~\cite{Pelaez:2003dy}).

The decays $f_0 / a_0(980) \to \gamma\ \rho(770) / \omega(782)$
are similar to decays $\phi \to \gamma\ a_0$ ($\gamma\ f_0$). The
interest to these processes was initiated in~\cite{Hanhart}.
Apparently they can be explained in terms of the same matrix
element (describing process with vector (V) and scalar (S)
particles, and photon in initial/final state) with $SU(3)$ flavor
modifications reflecting the type of vector particles. Decays $S
\to \gamma V$ may be studied experimentally in J\"ulich (with ANKE
and WASA at COSY)~\cite{COSY} and possibly in
Frascati~\cite{Ambrosino:2006gk}(with KLOE at DA$\Phi$NE or its
upgrade) and at BES.

Various phenomenological
models~\cite{Achasov_Ivanchenko,Harada06,Lemmer:2007qp,Hanhart:KKMol-phi,Hanhart:KKMol-gg,Close93,Vijande05,Black}
have been applied to calculation of these decays. At the same time
a consistent description in framework of Chiral Perturbation
Theory (ChPT) with pre-existing vector and axial-vector
mesons~\cite{EckerNP321} is lacking. This is an effective theory
of the strong and electromagnetic interactions at energies below
$1\ \text{GeV}$ and has symmetries of the underlying QCD. Strictly
speaking ChPT is an expansion in series $p^2/\Lambda_\chi^2, \
m^2/\Lambda_\chi^2 $, where $p$ is momentum, $m$ is mass of
pseudoscalar mesons, and chiral symmetry breaking scale
$\Lambda_\chi$ is of order $1~\text{GeV}$. Thus, formally, the
range of energies for scalar mesons is on the border of ChPT
applicability. Nevertheless, it is clear that suitable effective
Lagrangian for scalar mesons has to have much in common with ChPT
Lagrangian, because the coupling structures are guided by the
chiral symmetry. There are many successful applications of this
theory at energies about $1~\text{GeV}$ that make a useful
background for employing it in present problem.

In general, ChPT does not specify internal structure of
interacting particles. The model~\cite{EckerNP321} only assumes
that the scalar fields belong to $SU(3)$ flavor octet and singlet.
This Lagrangian is written down in Appendices~\ref{App:A}
and~\ref{App:B}, in particular, $L^A$ describes interaction of
pseudoscalar and vector mesons, and $L^B$ -- interaction of scalar
mesons with pseudoscalars. We test the singlet-octet mixing scheme
for the lightest scalar meson nonet
\bgea \label{eq:multiplet_sc}
\left\{
\begin{aligned}
a_0 =& S_3
,\\
f_0 =& S^{sing}\, \cos \theta - S_8\, \sin \theta
,\\
\sigma =& S^{sing}\, \sin \theta + S_8\, \cos \theta ,
\end{aligned}
\right.
 \enea
where $S_3$ is the neutral isospin-one, $S_8$ is isospin-zero
members of flavor octet and $S^{sing}$ is flavor singlet. $\theta$
is the octet-singlet mixing angle, and $\sigma = f_0(600)$. In
particular we are interested in whether $a_0(980)$ and $f_0(980)$
suit for members of this nonet. In principle this may not be the
case (see for example argumentation in~\cite{Klempt}) and
therefore should be verified. Radiative decays may help to clarify
this important issue.

The present paper considers decays $S \to \gamma \gamma$,
$\phi(1020) \to \gamma S$ and $S \to \gamma V$. We suppose that
the underlying dynamics of all above decays has much in common,
namely that the loops with pseudoscalar mesons form the dominant
mechanism. This assumption is consistently implemented in
Lagrangian~\cite{EckerNP321} and Section~\ref{sec:gen:approach}
presents calculation of decay amplitudes. On this way we prove
cancelation of divergences and gauge invariance of the amplitudes.
Along the calculations we use the dimensional regularization
method, see Appendix~\ref{App:D} for a brief overview of the
method and list of basic formulae. Some details of loop integrals
calculation and their analysis are also presented in
Section~\ref{sec:gen:approach}.

The are six coupling constants in Lagrangian ($F_V$, $G_V$, $c_d$,
$c_m$, $\tilde{c}_d$, $\tilde{c}_m$), and estimation of their
values is carried out in Section~\ref{sec:results_discussion}.
Under assumption of the resonance saturation the coupling
constants may be expressed in terms of chiral
LEC's~\cite{EckerNP321}. Available experimental data provides
certain constraints on these couplings.

After fixing the parameters we calculate the widths of various
decays with light scalar mesons and compare them with available
data and predictions of other models
(Section~\ref{sec:results_discussion}). We compare pion and kaon
loop contributions to the decays with $f_0$ meson in initial/final
state and demonstrate the importance of pion loops in $f_0 \to
\gamma\rho$ and $f_0 \to \gamma\gamma$ decays.

The virtual photon case, which is important for further
applications of the present model, is outlined in
Appendix~\ref{App:C}.


\section{Formalism for radiative decays amplitudes}
\label{sec:gen:approach}

\subsection{One-loop diagrams and chiral counting}
\label{sec:generalchiralpower}

\begin{figure*}
\begin{center}
\resizebox{0.85\textwidth}{!}{
  \includegraphics{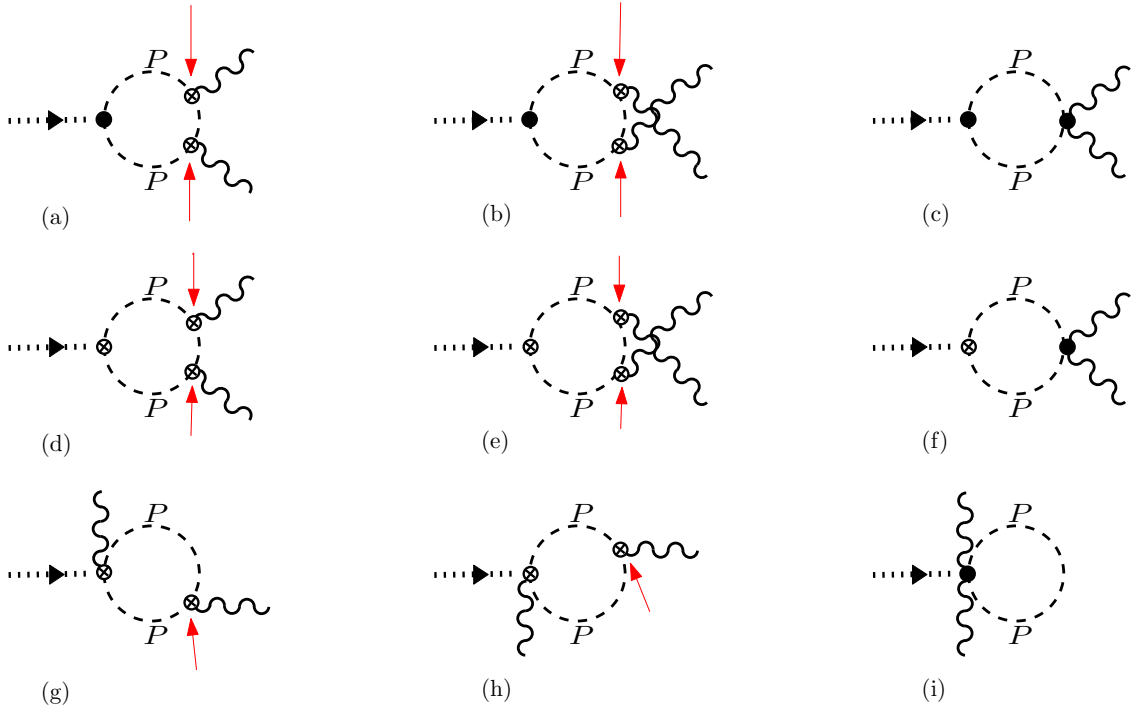}
}
\end{center}
\caption{Diagrams for decay of scalar meson (dotted line) into two
photons (wavy lines). Pseudoscalar mesons (dashed lines) in the loops
are: $(K^+K^-)$ for $a_0(980)$ decays, and $(K^+K^-),\,(\pi^+\pi^-)$ for
$f_0(980)$ decays. Solid (non-derivative coupling) and crossed
(derivative coupling) blobs represent $\mathcal{O}(p^2)$ vertices.
  Arrows mark the places, where form factors of pion and kaon would
arise for virtual photons (see discussion in Appendix \ref{App:C}).
   } \label{fig:p1}
\end{figure*}

From Lagrangian terms (\ref{eq:La}) and (\ref{eq:Lb}) one obtains
the sets of one-loop diagrams shown in
Figs.~\ref{fig:p1},~\ref{fig:p2} and~\ref{pic:p3}. In the present
approach we have no tree-level diagrams for the radiative
processes. Therefore the lowest-order amplitudes consist of
one-loop diagrams. The corresponding set of diagrams for $a_0/f_0
\to\gamma\gamma$ decay with pseudoscalar meson in the loop is
shown in Fig.~\ref{fig:p1}. This set of diagrams is complete since
it is obtained from Lagrangian which carries the chiral power not
less than the chiral power of any diagram.

\nopagebreak

The following rules~\cite{ChiralCounting} are used to count the
chiral power of any diagram. These counting rules provide one with
a guiding idea of which diagrams should be included and which
should not, when forming a set of relevant diagrams at any given
order.  Pseudoscalar fields $\Phi$, scalar fields $S$ and vector
fields (in tensor representation) $\rho_{\mu\nu}$,
$\omega_{\mu\nu}$, $\phi_{\mu\nu}$ carry zero chiral
power~$\mathcal{O}(p^0)$; derivative or external source (like
electromagnetic field $B_\mu$) has unit chiral
power~$\mathcal{O}(p)$; pseudoscalar-meson mass ($m_\pi$, $m_K$)
also carries unit power~$\mathcal{O}(p)$ (so that the mass matrix
$\chi$ in Appendix~\ref{App:A} is $\mathcal{O}(p^2)$). The
propagator of pseudoscalar meson is counted as
$\mathcal{O}(p^{-2})$.

All coupling constants in Lagrangian ($c_d$, $c_m$, $\tilde{c}_d$,
$\tilde{c}_d$, $F_V$, $G_V$) are $\mathcal{O}(p^{0})$, and the
power of any vertex is determined only by the structure of the
corresponding term in Lagrangian. In addition, the loop
integration over particle momentum adds $\mathcal{O}(p^{4})$.
Applying these rules one can show that each diagram in
Fig.~\ref{fig:p1},~\ref{fig:p2} and~\ref{pic:p3} has chiral power
$\mathcal{O}(p^4)$.


\subsection{Radiative decays \lowercase{$a_0/f_0~\to~\gamma\gamma$}}
\label{sec:rd1}

Consider the amplitude of $a_0/f_0 \to \gamma\gamma$ decay. Let scalar
meson have 4-momentum $p$, photons have polarization vectors
$\epsilon^{(1)}_\mu$ and $\epsilon^{(2)}_\nu$, and 4-momenta $q_{1}^\mu$
and $q_{2}^\nu$. Suppose that a positive charge runs clockwise in the
loop. First, we write down amplitude for the first three diagrams (a-c),
as this set of diagrams is often used in various approaches for radiative
decays~\cite{Achasov_Ivanchenko,Harada06,Hanhart}.

The invariant amplitude $\mathcal{M}_{abc}$ corresponding to diagrams
(a-c) in Fig.~\ref{fig:p1} is expressed through the tensor
$T^{\mu\nu}_{S(P)\gamma\gamma}$:
 \bgea -\imath \mathcal{M}_{abc} &=&
\epsilon^{(1)\ast}_\mu \ \epsilon^{(2)\ast}_\nu \ T^{\mu\nu}_{S(P)\gamma\gamma} .
 \enea
The indices $S(P)$ indicate that the scalar meson of type $S$ decays in
two photons via loop consisting of two intermediate pseudoscalar mesons
of type $P$ with mass $m_P$, i.e. $S(P) = \{a_0(KK), f_0(KK), f_0(\pi\pi)\}$.

Explicitly we find
 \bgea \label{eq:amplitude:nonder1}
T^{\mu\nu}_{S(P)\gamma\gamma} &=& -\frac{e^2 g_{SPP}}{f_\pi^2} \Bigl\{ 2
g^{\mu\nu} \int \! \! \frac{d^4l}{(2\pi)^4} \;\Delta_l \Delta_{l-p}
\nn\\&& + I^{\nu\mu}(q_1,q_2) + I^{\mu\nu}(q_2,q_1) \Bigr\} ,
\\
I^{\nu\mu}(q_1,q_2) &\equiv& \int  \! \!  \frac{d^4l}{(2\pi)^4} \;
(2l-p-q_1)^\nu (2l-q_1)^\mu \;
\nn\\&&
\times\; \imath \Delta_l \Delta_{l-p} \Delta_{l-q_1}
,
\nn
 \enea
where $\Delta_{l} \equiv \imath (l^2 - m_P^2)^{-1}$. For $g_{SPP}$'s
we refer to Lagrangian~(\ref{eq:Lb}) in Appendix~\ref{App:B}.
Changing the integration variable $l^\prime = p - l$ (and $l^\prime = p +
q_1 - l$) and assuming that the possible divergence of the integrals is
not higher than the logarithmic, one can prove the gauge invariance of
the amplitude:
 \bgea
 q_{1 \mu}T^{\mu\nu}_{S(P)\gamma\gamma} =
q_{2  \nu}T^{\mu\nu}_{S(P)\gamma\gamma} = 0 .
 \label{eq:gauge_invariance}
 \enea
Making use of the change $l^\prime = p - l$ we deduce a useful
relation
 \bgea I^{\nu\mu}(q_1,q_2) = I^{\mu\nu}(q_2,q_1)
 \enea
which is connected with Bose-symmetry of the final photons.

By means of the Feynman parametrization and dimensional regularization
method (see~Appendix~\ref{App:D}) the
expression~(\ref{eq:amplitude:nonder1}) is reduced to
 \bgea
\label{eq:amplitude:nonder2} T^{\mu\nu}_{S(P)\gamma\gamma} &=& \frac{-2
\imath e^2}{(4\pi)^2} \frac{g_{SPP}}{f_\pi^2} \Bigl\{ g^{\mu\nu} \int_0^1
\! dx \ln [m_P^2 - p^2 x(1-x)] \nn\\ && - 2 g^{\mu\nu} \iint_0^1 x dx dy
\ln [C(x,y;q_1,q_2)] \nn\\&& - \iint_0^1 x dx dy
\frac{A^{\nu\mu}(x,y;q_1,q_2)}{C(x,y;q_1,q_2)} \Bigr\} , \enea \bgea
A^{\nu\mu}(x,y;q_1,q_2) &=& (q_1[2x(1-y)-2] + 2xy\ p - q_2)^\nu \nn\\&&
\times\; (q_1[2x(1-y)-1] + 2xy\ p)^\mu
\nn,\\
C(x,y;q_1,q_2) &=& q_1^2\ x(x-1)(1-y) + p^2 \ xy (x-1) \nn\\&& - q_2^2\
x^2 y (1-y) + m^2 \nn.
 \enea
The divergent parts of diagrams (a-c) in Fig.~\ref{fig:p1} cancel
each other.

For real photons in question
 \bgea
&q_1^2 = q_2^2 = 0 ,\nn\\
&\epsilon^{(1)}_\mu \;q_1^\mu = \epsilon^{(2)}_\mu \;q_2^\mu =0
\nn.
 \enea
These conditions simplify equation (\ref{eq:amplitude:nonder2}) to
 \bgea T^{\mu\nu}_{S(P)\gamma\gamma} &=& \frac{-4 \imath e^2}{(4\pi)^2}
\frac{g_{SPP}}{f_\pi^2} \bigl( g^{\mu\nu} - \frac{q_1^\nu
q_2^\mu}{q_1\!\cdot\!q_2} \bigr) \Psi(m_P^2; p^2; 0) .
 \enea
Here we define
 \bgea \label{eq:psi_defin1} \Psi(m_P^2; p^2; 0) \equiv
\iint_0^1 x dx dy \Bigl[1 + \frac{m_P^2}{p^2 xy(x-1)} \Bigr]^{-1}
\\
= \frac{1}{2} - \frac{m_P^2}{p^2} \int_0^1 \! \frac{dx}{x-1} \ln[1
+ x(x-1)\frac{p^2}{m_P^2}] ,\nn
\enea
\bgea Re\ \Psi(m_P^2; p^2;
0) &=& \! \frac{1}{2} \! - \! \frac{m_P^2}{p^2} \! \int_0^1 \! \!
\! \frac{dx}{x-1} \ln \left| \! 1 + x(x-1)\frac{p^2}{m_P^2} \!
\right|
,\nn\\
Im\ \Psi(m_\pi^2; p^2; 0) &=& \pi \frac{m_\pi^2}{p^2} \ln \Bigl|
\frac{1 + \sqrt{1 - 4 \frac{m_\pi^2}{p^2} } } {1 - \sqrt{1 - 4
\frac{m_\pi^2}{p^2} } } \Bigr|,
\nn\\
Im\ \Psi(m_K^2; p^2; 0) &=& 0 \nn. \enea The integrals
$\Psi(m_P^2; p^2; 0)$ are calculated numerically and presented in
Table~\ref{table:psi}. The scalar-meson invariant mass
$(p^2)^{1/2}$ is equal to $M_s$ -- the mass of $f_0$, $a_0$ (and
$\sigma$ for completeness), while $m_P$ is equal to the mass of
pseudoscalar meson $\pi$ or $K$ in the loop \footnote{When working
with integrals~(\ref{eq:psi_defin1}) it is convenient to use the
identity $\int_0^1 (1-2x) f(y) dx = 0$ for any function $f(y)$,
where $y=x(1-x)$.}.

\begin{table}
\caption{Values of the loop integrals. Assumed physical values of scalar meson masses: $M_{f_0}=980~\text{MeV}$ and $M_{a_0}=984.7~\text{MeV}$
}
\label{table:psi}
\begin{center}
\begin{tabular}{rcl}
\hline\noalign{\smallskip} $\Psi(m_K^2; M_{a_0}^2; M_\phi^2)$&=&
$0.0749 + 0.244 \, \imath$
\\
$\Psi(m_K^2; M_{f_0}^2; M_\phi^2)$&=& $0.1295 + 0.216 \, \imath$
\\
$\Psi(m_\pi^2; M_{a_0}^2 ; 0)$&=& $0.5510 - 0.244 \, \imath $
\\
$\Psi(m_K^2; M_{a_0}^2 ; 0)$&=&
$-0.63$
\\
$\Psi(m_\pi^2; M_{f_0}^2 ; 0)$ &=& $0.5507 - 0.246 \, \imath$
\\
$\Psi(m_K^2; M_{f_0}^2 ; 0)$&=&
$-0.57$
\\
$\Psi(m_\pi^2; M_{\sigma}^2 ; 0)$ &=& $0.3545 - 0.5664 \, \imath$
\\
$\Psi(m_K^2; M_{\sigma}^2 ; 0)$ &=&
$-0.0529 $
\\
$\Psi(m_\pi^2; M_{f_0}^2; M_\rho^2)$&=& $0.128 - 0.0169 \, \imath$
\\
$\Psi(m_K^2; M_{a_0}^2; M_\rho^2)$&=&
$-0.4048$
\\
$\Psi(m_K^2; M_{a_0}^2; M_\omega^2)$&=&
$-0.3988$
\\
$\Psi(m_K^2; M_{f_0}^2; M_\rho^2)$&=&
$-0.3466$
\\
$\Psi(m_K^2; M_{f_0}^2; M_\omega^2)$&=&
$-0.3407$
\\
\noalign{\smallskip}\hline
\end{tabular}
\end{center}
\end{table}

Now we consider diagrams (d-i) in Fig.~\ref{fig:p1}. At first glance,
these diagrams, which include derivative coupling for scalar mesons
\footnote{We also include here diagram (i), though it has no derivative
coupling in $SPP\gamma\gamma$ vertex. This is convenient due to its
cancelation with contribution of diagram~(f). }, are more complicated due
to momentum dependence of the $S P P $ vertex. In fact these diagrams can
be treated similarly to the previous case. To demonstrate this let us
define
 \bgea \label{eq:Md_i} -\imath \mathcal{M}_{d-i} &=&
\epsilon^{(1)}_\mu \epsilon^{(2)}_\nu
\hat{T}^{\mu\nu}_{S(P)\gamma\gamma},
 \enea
where symbol ``hat'' is used hereafter to indicate the derivative
coupling.
Next use the identity
 \bge
 l(l-p)\Delta_l \Delta_{l-p} =
\! \frac{\imath}{2} \! \left(
    \Delta_l \!+\! \Delta_{l-p} + \! \imath (p^2 \! - \! 2 m_P^2)\Delta_l \Delta_{l-p}
\right), \nn
 \ene
 and change integration variables as above in order to combine six terms
in $\mathcal{M}_{d-i}$ in such a way that the contribution of
diagram (i) cancels the contribution of (f), and diagrams (g), (h)
are cancelled by part of (d) and part of (e). In this way the
derivative coupling amplitude
$\hat{T}^{\mu\nu}_{S(P)\gamma\gamma}$ is related to the
non-derivative coupling amplitude $T^{\mu\nu}_{S(P)\gamma\gamma}$
 \bgea
 \epsilon^{(1)}_\mu \epsilon^{(2)}_\nu \
 \hat{T}^{\mu\nu}_{S(P)\gamma\gamma} &=&
\frac{\hat{g}_{SPP}}{g_{SPP}}(m_P^2 - p^2/2) \
\epsilon^{(1)}_\mu \epsilon^{(2)}_\nu \
T^{\mu\nu}_{S(P)\gamma\gamma} \nn .
 \enea

Combining contributions of all diagrams in Fig.~\ref{fig:p1} one obtains
the total $\mathcal{O}(p^4)$ invariant amplitude
 \bgea
\label{eq:amplitudeA} - \! \imath \mathcal{M}_{a_0\to\gamma\gamma}
&=& \frac{-4\imath e^2}{(4\pi)^2 f_\pi^2}\ \Psi(m_K^2; p^2; 0)
A_K(p^2)
  \\
  && \times\;
\left( \epsilon^{(1)\ast}\cdot\epsilon^{(2)\ast} -
\frac{\epsilon^{(1)\ast}\cdot q_{2} \; \epsilon^{(2)\ast}\cdot
q_{1} }{q_{1}\cdot q_{2}} \right), \nn
  \\
  \label{eq:amplitudeF}
- \! \imath \mathcal{M}_{f_0\to\gamma\gamma} &=& \frac{-4\imath
e^2}{(4\pi)^2 f_\pi^2}\ [ B_\pi(p^2) \Psi(m_\pi^2; p^2; 0)
  \\
  && \;
+ B_K(p^2) \Psi(m_K^2; p^2; 0) ] \nn
  \\
  && \times\;
\left( \epsilon^{(1)\ast}\cdot\epsilon^{(2)\ast} -
\frac{\epsilon^{(1)\ast}\cdot q_{2} \; \epsilon^{(2)\ast}\cdot
q_{1} }{q_{1}\cdot q_{2}} \right), \nn
 \enea
where
 \bgea
 \label{eq:four_coef}
    A_K(p^2) &\equiv & \hat{g}_{aKK} (m_K^2 - p^2/2) + g_{aKK}, \nn\\
    B_K(p^2) &\equiv & \hat{g}_{fKK} (m_K^2 - p^2/2) + g_{fKK}, \nn\\
    B_\pi(p^2) &\equiv & \hat{g}_{f\pi\pi} (m_\pi^2 - p^2/2) + g_{f\pi\pi}.
 \enea


\subsection{Radiative decays \lowercase{$\phi(1020) \to \gamma~a_0/f_0$}}
\label{sec:rd2}

\begin{figure*}
\begin{center}
\resizebox{0.85\textwidth}{!}{
  \includegraphics{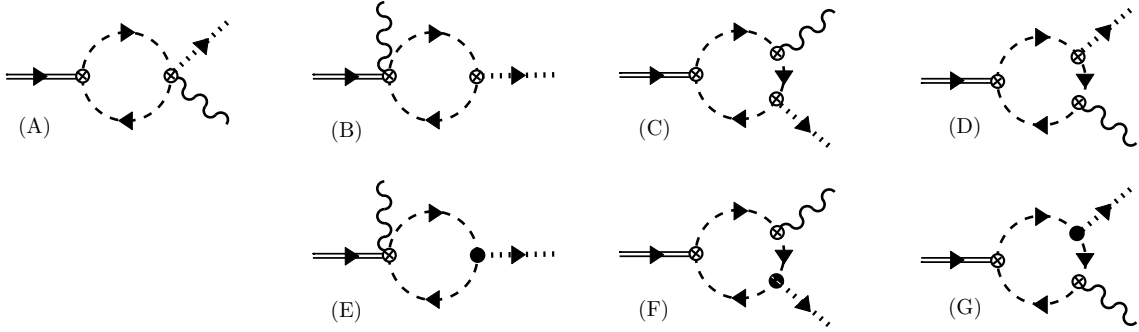}
}
\end{center}
\caption{Diagrams for $\phi(1020)$ to $\gamma a_0(980)$ or $\gamma
f_0(980)$ decays. Pseudoscalar mesons here are: $(K^+K^-)$. Solid and
crossed blobs stand for $\mathcal{O}(p^2)$ vertices, the latter indicate
derivative coupling terms. } \label{fig:p2}
\end{figure*}

Let vector meson $\phi(1020)$ with momentum  $Q$ decay into scalar
meson $a_0(980)$ (or $f_0(980)$) with momentum $p$ and photon with
momentum $q$, i.e. $\phi(Q) \to \gamma (q) + a_0/f_0(p)$. Diagrams
corresponding to these reactions are shown in Fig.~\ref{fig:p2}.

Let polarization vector for the $\phi$-meson be $E_\mu$, and that
for the photon $\epsilon_\nu$. Apparently $q^\nu \epsilon_\nu=0$,
\ $Q^\mu E_\mu=0$. We describe vector meson $\phi(1020)$ by
antisymmetric tensor field carrying the indices $\mu \lambda$.
Thus we employ the normalization for one-particle matrix
element~\cite{EckerNP321} \bge \langle 0\, |\, \phi_{\mu \lambda}
(0) \,|\, \phi, Q \rangle = \imath\, M_{\phi}^{-1} [Q_\mu
E_\lambda - Q_\lambda E_\mu]. \ene
 The invariant amplitude reads: \bgea -\imath
\mathcal{M}_{\phi \to \gamma S} &=& \epsilon_\nu^\ast \  \imath \frac{Q_\mu
E_\lambda - Q_\lambda E_\mu}{M_\phi}
\nn\\
&&\times \ \Bigl( T_{\phi \to \gamma S}^{\mu\lambda\nu} + \hat{T}_{\phi
\to \gamma S}^{\mu\lambda\nu} \Bigr) ,
 \enea
where tensor with ``hat'' for diagrams (A-D) is
 \bgea
\label{eq:deriv_g} \hat{T}_{\phi\to \gamma S}^{\mu\lambda\nu} &=&
\hat{T}_A^{\mu\lambda\nu} + \hat{T}_B^{\mu\lambda\nu}
+\hat{T}_C^{\mu\lambda\nu} +\hat{T}_D^{\mu\lambda\nu}
  , \nn \\
\hat{T}_A^{\mu\lambda\nu} &=&
\frac{-\imath e G_V \hat{g}_{SKK}}{\sqrt{2} f_\pi^4}\int \frac{d^4l}{(2\pi)^4}
\bigl( Q^\mu l^\lambda - (\mu \leftrightarrow \lambda) \bigr)
\nn\\&& \times \;
\bigl( 2l - Q \bigr)^\nu \Delta_l \Delta_{l-Q}
  , \nn \\
\hat{T}_B^{\mu\lambda\nu} &=& \frac{-\imath e
\hat{g}_{SKK}}{\sqrt{2} f_\pi^4}\int \frac{d^4l}{(2\pi)^4} \nn \\
&& \times \Bigl[ \bigl( G_V\ Q^\mu + \frac{1}{2} (F_V - 2\ G_V)
q^\mu \bigr) g^{\nu\lambda} \nn \\ && - \left( \mu \leftrightarrow
\lambda \right) \Bigr ]
 \;
 l\bigl( l - Q + q \bigr) \Delta_l \Delta_{l-Q + q}
  , \nn \\
\hat{T}_C^{\mu\lambda\nu} &=& \hat{T}_D^{\mu\lambda\nu}  \nn \\
&=& \frac{e G_V \hat{g}_{SKK}}{\sqrt{2} f_\pi^4}\int
\frac{d^4l}{(2\pi)^4} \bigl( Q^\mu l^\lambda - (\mu
\leftrightarrow \lambda) \bigr) \nn\\&& \times \; \bigl( 2l - q
\bigr)^\nu (l-q)(l-Q)\; \Delta_l \Delta_{l-q}\Delta_{l-Q} ,
 \enea
and tensor without ``hat'' for diagrams (E-G) reads
 \bgea
 \label{eq:non-deriv_g} T_{\phi \to \gamma
S}^{\mu\lambda\nu} &=& T_E^{\mu\lambda\nu} +
T_F^{\mu\lambda\nu}+T_G^{\mu\lambda\nu}
    , \nn\\
T_E^{\mu\lambda\nu} &=& \frac{-\imath e  g_{SKK}}{\sqrt{2}
f_\pi^4} \int \frac{d^4l}{(2\pi)^4} \Delta_l \Delta_{l-Q+q}
 \Bigl[ g^{\nu\lambda} \bigl( G_V\ Q^\mu
 \nn\\ && \; +
\frac{1}{2} (F_V - 2\ G_V) q^\mu \bigr) - \left( \mu
\leftrightarrow \lambda \right) \Bigr]
    , \nn\\
T_F^{\mu\lambda\nu} &=& T_G^{\mu\lambda\nu} \nn\\
&=& \frac{- e G_V g_{SKK}}{\sqrt{2} f_\pi^4} \int
\frac{d^4l}{(2\pi)^4} \Delta_l \Delta_{l-q} \Delta_{l-Q} \bigl( 2l
- q \bigr)^\nu \nn\\&& \times \; \bigl( Q^\lambda l^\mu - (\mu
\leftrightarrow \lambda) \bigr)
.
 \enea

The consideration shows that divergent parts of the
amplitudes~(\ref{eq:deriv_g}) and~(\ref{eq:non-deriv_g}) which do
not cancel are proportional to $(F_V - 2\ G_V)(g^{\nu\lambda}q^\mu
- g^{\nu\mu}q^\lambda)$. Therefore we employ the relation
 \begin{equation}F_V = 2\ G_V
 \label{eq:FV-GV}
  \end{equation}
between electromagnetic and strong coupling constants of vector
mesons (see Appendix~\ref{App:A}) in order to make the amplitudes
finite. Actually this relation does not follow from the chiral
symmetry. However it naturally appears in alternative approaches,
Hidden Local Gauge Symmetry Model~\cite{HGS} and massive
Yang-Mills theory~\cite{MYM}. This aspect has been addressed
in~\cite{EckerPLB223}.  In Appendix~\ref{App:A} we discuss
accuracy of (\ref{eq:FV-GV}) based on experiment.

Making use of~(\ref{eq:FV-GV}) and identities
 \bgea \label{eq:totozh} \imath
(\Delta_{l-q} + \Delta_{l-Q})&=& \Delta_{l-q}\Delta_{l-Q}\! \left[
\!p^2\! -\! 2 m_K^2\! +\! 2 (l-q)(l-Q)\! \right]
, \nn\\
\imath (\Delta_{l} + \Delta_{l-Q+q})&=&\!
\Delta_{l}\Delta_{l-Q+q}\! \left[ p^2 - 2 m_K^2 + 2 l(l-Q+q)
\right] \nn,
 \enea
 in (\ref{eq:deriv_g}) one can prove that
 \bgea \label{eq:equal}
\left(
\hat{T}_{\phi\to \gamma S}^{\mu\lambda\nu} -
\frac{\hat{g}_{SKK}}{{g}_{SKK}} \bigl( m_K^2 - \frac{p^2}{2} \bigl)
T_{\phi\to \gamma S}^{\mu\lambda\nu}
\right)
&&\nn \\
\times
\epsilon_\nu^\ast \  \imath \frac{Q_\mu E_\lambda - Q_\lambda
E_\mu}{M_\phi}
&=&0.
 \enea
Finally
 \bgea
 \label{eq:inv_amp_gv}
 -\imath \mathcal{M}_{\phi \to \gamma S} &=&
 \Bigl( 1 + \frac{\hat{g}_{SKK}}{{g}_{SKK}} \bigl( m_K^2 -
\frac{p^2}{2} \bigl) \Bigr)
\\&& \times \;
T_{\phi\to \gamma S}^{\mu\lambda\nu} \epsilon_\nu^\ast \ \imath \frac{Q_\mu
E_\lambda - Q_\lambda E_\mu}{M_\phi} \nn.
 \enea

In calculation of the amplitude the Feynman parametrization and
dimensional regularization method are applied (Appendix~\ref{App:D}).
Then eq.~(\ref{eq:inv_amp_gv}), with the use of (\ref{eq:non-deriv_g}),
reads
 \bgea
\label{eq:amp_via_int} -\imath \mathcal{M}_{\phi \to \gamma S} &=&
\frac{- \imath e G_V g_{SKK} Q^2}{\sqrt{2} f_\pi^4 (4\pi)^2 M_\phi}
\\&&\nn \times\, \Bigl( 1 +
\frac{\hat{g}_{SKK}}{{g}_{SKK}} \bigl( m_K^2 - \frac{p^2}{2}
\bigl) \Bigr)
\\&&\nn \times\,
\Bigl [ 4(Q\cdot\epsilon^\ast)(q\cdot E) I_1 - (\epsilon^\ast
\cdot E)(I_2 - 2 I_3) \Bigr ] , \enea where \bgea I_2 &=& \int_0^1
dx \ln (m_K^2 - p^2x(1-x))
,\nn\\
I_3 &=& \iint_0^1 x dx dy \ln (m_K^2 - Q^2x(1-x) + 2 xy(1-x)Q\cdot q)
,\nn\\
I_1 &=& \iint_0^1 x dx dy \frac{xy(1-x)}{m_K^2 - Q^2x(1-x) + 2 xy(1-x)Q\cdot q}
\nn\\
&=& \frac{1}{4 Q\cdot q} \bigl( I_2 - 2 I_3 \bigr),
\label{eq:int_I_1}
 \enea
\bgea
\label{eq:psidef2}
I_2 - 2 I_3&=&1 - \int_0^1 dx
\frac{m_K^2 - M_\phi^2 x(1-x)}{(M_\phi^2 - p^2)x(1-x)}
\\ && \nn \times
\ln{\frac{m_K^2 - p^2 x(1-x)}{m_K^2 - M_\phi^2 x(1-x)}}
 \equiv 2 \ \Psi(m_K^2, p^2, M_\phi^2).
 \enea
In terms of $\Psi(m_K^2, p^2, M_\phi^2)$ the invariant
amplitude~(\ref{eq:amp_via_int}) reads
 \bgea
\label{eq:rd2:2} -\imath \mathcal{M}_{\phi \to \gamma S} &=& \imath
\frac{\sqrt{2}e M_\phi G_V}{f_\pi^4 (4\pi)^2} \, 2 \Psi(m_K^2,
p^2, M_\phi^2)
\\
&& \times \Bigl[ \epsilon^\ast \cdot E - \frac{1}{Q \!\cdot\! q} (
Q\cdot \epsilon^\ast)\;(q\cdot E^\ast) \Bigr]
\nn
  \\
  && \times\;
  \left[
\begin{aligned}
A_K(p^2)   \;\;\;\;\; & \text{for}\;\;\;\; \phi \to KK \to a_0\gamma\\
B_K(p^2)   \;\;\;\;\; & \text{for}\;\;\;\; \phi \to KK \to f_0 \gamma
\end{aligned}
\right.
\nn
 .
 \enea

Compare definition (\ref{eq:psidef2}) with that of $\Psi(m_P^2,
p^2, 0)$ in~(\ref{eq:psi_defin1}). Real and imaginary parts of
$\Psi(m_K^2, p^2, M_\phi^2)$ at $p^2 = M_s^2$ are
 \bgea
 Re \;\Psi(m_K^2,
M_s^2, M_\phi^2) &=& \frac{1}{2} - \frac{1}{2} \int_0^1 dx \frac{m_K^2 -
M_\phi^2 x(1-x)}{(M_\phi^2 - M_s^2)x(1-x)}
\nn\\
&& \times \; \ln \Bigl| {\frac{m_K^2 - M_s^2 x(1-x)}{m_K^2 - M_\phi^2
x(1-x)}} \Bigr| ,\nn
\\
Im \;\Psi(m_K^2, M_s^2, M_\phi^2) &=& \frac{\pi M_\phi^2}{M_\phi^2-M_s^2}
\Biggl( \sqrt{\frac{1}{4} - \frac{M_K^2}{M_\phi^2}}
\nn\\
&& +\frac{M_K^2}{M_\phi^2} \ln \Biggl| \frac{1- \sqrt{1-
4\frac{M_K^2}{M_\phi^2}}}{1 + \sqrt{1- 4\frac{M_K^2}{M_\phi^2}}}
\Biggr| \Biggr). \label{eq:rd2:4}
 \enea
Numerical calculation of $\Psi(m_K^2, M_s^2, M_\phi^2)$ leads to
values shown in Table~\ref{table:psi} (see
also~\cite{Achasov_Ivanchenko} for analytic expression of the
integral (\ref{eq:psidef2})).


\subsection{Radiative decays \lowercase{$f_0/ a_0 \to \gamma~\rho/\omega$}}
\label{sec:rd3}

\begin{figure*}
\begin{center}
\resizebox{0.85\textwidth}{!}{
  \includegraphics{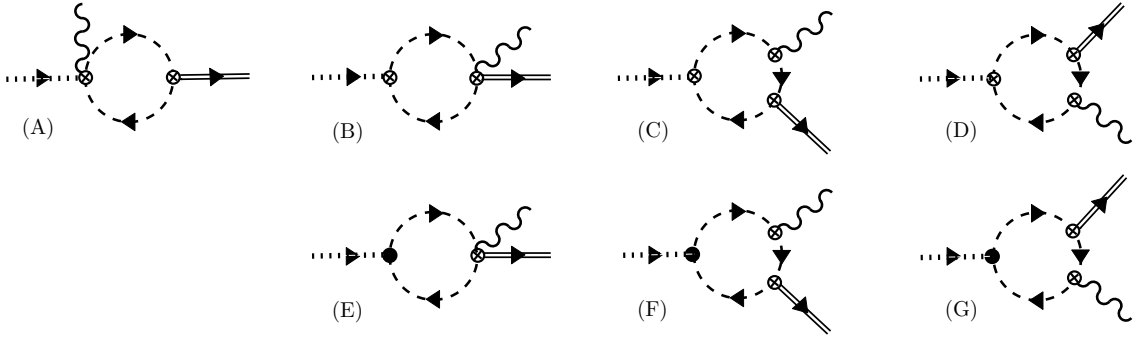}
}
\end{center}
\caption{ Diagrams for $f_0(980)/ a_0(980)\to\gamma
\rho(770)/\omega(782)$ decays. Pseudoscalar mesons in the loops are:
$(K^+K^-)$ for all decays, and $(\pi^+ \pi^-,\ K^+ K^-)$ for $f_0(980)
\to \gamma \rho(770)$. Solid and crossed blobs stand for
$\mathcal{O}(p^2)$ vertices, the latter are derivative coupling terms. }
\label{pic:p3}
\end{figure*}

Decay of scalar meson into vector meson with radiation of photon ($f_0 /
a_0 \to \gamma \ \rho /\omega$) in the lowest order is represented by
$\mathcal{O}(p^4)$ diagrams shown in Fig.~\ref{pic:p3}. The vertices
follow from $L^A$ in (\ref{eq:La}) and $L^B$ in (\ref{eq:Lb}). The
structure of the matrix element for these decays is very similar to that
in (\ref{eq:rd2:2}). One can replace $\phi(1020)$ by $\rho(770)$ (or
$\omega(782)$), take into account flavor $SU(3)$ factor in the $VPP$
vertices, and select the pseudoscalar particles in the loops allowed by
symmetries of the strong interaction. For relevant $SU(3)$ relations
see Appendix~\ref{App:A}. Taking
into account that $a_0(980)$ and $\omega (782)$ do not couple to two
pions, one is left with $K^+K^-$ loop for $a_0 \to \gamma V$ and $f_0 \to
\gamma \omega$ decays, and both $\pi^+\pi^-$ and $K^+K^-$ loops for $f_0
\to \gamma \rho$ decay.

The matrix elements read
 \bgea
\label{M:aToGammaV} -\imath \mathcal{M}_{a_0 \to \gamma V} &=&
\frac{-\imath e M_V G_V }{ f_\pi^4 (4 \pi)^2} \Bigl[ \epsilon^\ast \cdot
E^\ast - \frac{1}{Q \!\cdot\! q} (Q \cdot \epsilon^\ast) (q\cdot
E^\ast) \Bigr]
\nn\\
 && \times\ A_K(p^2)\ \Psi(m_K^2; p^2; M_V^2) ,
  \enea
  \bgea
\label{M:fToGammaRho}
 -\imath \mathcal{M}_{f_0 \to \gamma \rho} &=&
\frac{-\imath e M_\rho G_V}{ f_\pi^4 (4 \pi)^2} \Bigl[ \epsilon^\ast
\cdot E^\ast - \frac{1}{Q \!\cdot\! q} (Q\cdot \epsilon^\ast)
(q\cdot E^\ast)
\Bigr] \nn\\
 && \times\ \bigl( B_K(p^2) \Psi(m_K^2; p^2; M_\rho^2)
\nn\\
 && + 2 B_\pi(p^2) \Psi(m_\pi^2; p^2; M_\rho^2) \bigl) ,
  \enea \bgea
\label{M:fToGammaOmega}
 -\imath \mathcal{M}_{f_0 \to \gamma \omega} &=&
\frac{- \imath e M_\omega G_V }{ f_\pi^4 (4 \pi)^2} \Bigl[ \epsilon^\ast
\cdot E^\ast - \frac{1}{Q \!\cdot\! q} (Q\cdot\epsilon^\ast)
(q\cdot E^\ast) \Bigr] \nn\\&& \times\ B_K(p^2)\ \Psi(m_K^2; p^2;
M_\omega^2) ,
 \enea
where notation for momenta and polarization vectors is the same as in
Section~\ref{sec:rd2}.

The loop integrals $\Psi(m_\pi^2; p^2; M_V^2)$ and $\Psi(m_K^2;
p^2; M_V^2)$ can also be defined by (\ref{eq:psidef2}). Their
numerical values are shown in Table~\ref{table:psi}.

To make a correspondence with results of~\cite{Close93} we can write
the loop integrals for $f_0/ a_0\to\gamma \ \rho/\omega$
diagrams in Fig.~\ref{pic:p3} in the form
 \bgea \label{eq:integrals_close_jue1} & Re\, \Psi(m_\pi^2;& M_{s}^2;
M_\rho^2) = \frac{1}{2} - \frac{1}{a-b} \Bigl( \ln^2 \frac{1 +
\sqrt{1-4/b}}{1 - \sqrt{1-4/b}} \nn\\&& - \ln^2 \frac{1 +
\sqrt{1-4/a}}{1 - \sqrt{1-4/a}} \Bigr) \nn\\&& + \frac{a}{2(a-b)}
\Bigl( \sqrt{1-4/b} \ln \frac{1 + \sqrt{1-4/b}}{1 - \sqrt{1-4/b}}
\nn\\&& - \sqrt{1-4/a} \ln \frac{1 + \sqrt{1-4/a}}{1 -
\sqrt{1-4/a}} \Bigr),  \enea  \bgea
 & Im\, \Psi(m_\pi^2;& M_{s}^2;
M_\rho^2) = \frac{\pi}{a-b} \Bigl( \ln \frac{1 + \sqrt{1-4/b}}{1 -
\sqrt{1-4/b}} \nn\\&& - \ln \frac{1 + \sqrt{1-4/a}}{1 -
\sqrt{1-4/a}} \Bigr)
\\&&
-
\frac{\pi a }{2(a-b)}
\Bigl(
\sqrt{1-4/b}
-
\sqrt{1-4/a}
\Bigr),
\nn
 \enea
with $a=M_\rho^2 / m_\pi^2$ and $b=M_s^2 / m_\pi^2$
for pions in the loop.
For kaons in the loop one finds
 \bgea
\label{eq:integrals_close_jue2} \Psi(m_K^2; M_{s}^2; M_V^2) &=&
\frac{1}{2} + \frac{1}{a-b} \Bigl( \arcsin^2 \frac{\sqrt{b}}{2} -
\arcsin^2 \frac{\sqrt{a}}{2} \Bigr) \nn\\&& + \frac{a}{a-b} \Bigl(
\sqrt{4/b-1} \arcsin^2 \frac{\sqrt{b}}{2} \nn\\&& - \sqrt{4/a-1}
\arcsin^2 \frac{\sqrt{a}}{2} \Bigr)
,
 \enea
 where $a=M_V^2 / m_K^2$ and $b=M_s^2 / m_K^2$.
Numerical values obtained from these analytic expressions agree
with those in Table~\ref{table:psi} obtained from direct numerical
integration. For the integrals (\ref{eq:psi_defin1}) and
(\ref{eq:psidef2}) one can also deduce analytic expressions
from~\cite{Close93} by choosing appropriate $a$ and $b$.

\section{Results and discussion}
\label{sec:results_discussion}

\begin{table*}
\caption{Particle properties (data from PDG~\cite{pdg}) which are needed
in calculations } \label{table:particles}
\begin{center}
\begin{tabular}{l|lccc}
\hline\noalign{\smallskip}
Particle& $I^G(J^{PC})$& Mass & Width & Major hadronic\\
&& ($\text{MeV}$) & ($\text{MeV}$) & decay channels \\
\noalign{\smallskip}\hline\noalign{\smallskip}
${a_0(980)}$ & $1^-(0^{++})$ & $984.7\pm 1.2$        & $50-100$ & $\pi \eta$ \\
${f_0}(980)$ & $0^+(0^{++})$ & $980 \pm 10 $         & $40-100$ & $\pi\pi$ \\
$\sigma={f_0}(600)$ & $0^+(0^{++})$~\cite{pdg} & $400\;-\;1000 $& $600-1000$ & $\pi\pi$ \\
${\sigma={f_0}(600)}$ & $0^+(0^{++})$~\cite{MURAMATSU_02} & $513 \pm 32$ & $335\pm 67$ & $\pi\pi$\\
$\pi^\pm$ & $1^-(0^-)$ & $139.57018 \pm 0.00035$     & mean life $2.6\ \times 10^{-8}\ s $ &  \\
$K^\pm$ & $1/2(0^-)$ & $493.677 \pm 0.016$           & mean life $1.24\ \times 10^{-8}\ s $ & $\pi^\pm \pi^0$ \\
${\phi}(1020)$ & $0^-(1^{--})$ & $1019\pm 0.019$     & $4.26$ &  $K^+K^-$, $K^0_L K^0_S$,
\\ &&&&
$\rho\pi + 3\pi$ \\
\noalign{\smallskip}\hline
\end{tabular}
\end{center}
\end{table*}

\subsection{Widths and estimates for chiral couplings}
\label{sec:generalcalculation} First of all there are direct
decays, which can be described from (\ref{eq:Lb}) at tree-level.
They represent the dominant channels: $a_0\to\pi\eta$ for
isotriplet and $f_0\to\pi\pi$ for isosinglet scalar mesons
 \bgea
\label{width:ape}
\Gamma_{a_0\to\pi\eta} &=& \frac{1}{8 \pi p^2} \sqrt{\frac{(p^2+m_\pi^2-m_\eta^2)^2}{4p^2}-m_\pi^2}
\frac{|A_{\pi\eta}(p^2)|^2}{f_\pi^4} , \\
\label{width:fpp} \Gamma_{f_0\to \pi\pi} &=& (1+\frac{1}{2}) \frac{1}{8
\pi p^2} \sqrt{p^2/4 - m_\pi^2} \frac{1}{f_\pi^4} |B_\pi(p^2)|^2,
 \enea
here $A_{\pi\eta}(p^2)$ is introduced by analogy with~(\ref{eq:four_coef})
 \bgea
A_{\pi\eta}(p^2) &\equiv & \hat{g}_{a\pi\eta} (m_\eta^2 + m_\pi^2 - p^2)/2 + g_{a\pi\eta},
 \enea
and the decays $a_0\to K\bar{K}$ and $f_0\to K\bar{K}$:
 \bgea
\label{width:akk}
    \Gamma_{a_0\to K\bar{K}} &=& 2 \frac{1}{8 \pi p^2} \sqrt{p^2/4 - m_K^2}
\frac{1}{f_\pi^4} |A_K(p^2)|^2, \\
\label{width:fkk}
    \Gamma_{f_0\to K\bar{K}} &=& 2 \frac{1}{8 \pi p^2} \sqrt{p^2/4 - m_K^2}
\frac{1}{f_\pi^4} |B_K(p^2)|^2.
 \enea
The invariant mass of scalar meson is $\sqrt{p^2}$.

For decays into $K\bar{K}$ in~(\ref{width:akk})
and~(\ref{width:fkk}) one includes factor $2$, (as $K\bar{K} =
K^+K^-, K^0\bar{K}^0$), and $(1+ 1/2\;(2\times1/2)^2)=3/2$ for
$\pi\pi$ in~(\ref{width:fpp}): 1 - from charged pions, $1/2$ -
from the identity of neutral pions, $(1/2)^2$ - because the
neutral pions interact two times weaker than the charged ones
($\stackrel{\rightarrow}{\pi}^2 = \pi^0\pi^0 + 2\pi^+\pi^-$), and
$2$ is the symmetry factor in the vertex with two identical
neutral pions.

The widths of our premium interest are built up of $A_K(p^2)$, $B_K(p^2)$
and $B_\pi(p^2)$~(\ref{eq:four_coef}), loop integrals $\Psi$ and
phase-space factors. Thus, through relations~(\ref{eq:four_coef}),
they depend on Lagrangian couplings $c_d, c_m, \tilde{c}_d,
\tilde{c}_m$ and singlet-octet mixing angle $\theta$ for scalar
mesons (see Appendix~\ref{App:B}).

The widths for $a_0\to\gamma\gamma$, $f_0\to\gamma\gamma$  decays read
 \bgea
 \label{width:agg}
\Gamma_{a_0\to\gamma\gamma} &=& \frac{1}{32\pi \sqrt{p^2}}
\frac{e^4}{8\pi^4 f_\pi^4 }
|A_K(p^2)\ \Psi(m_K^2; p^2; 0)|^2, \\
\label{width:fgg}
\Gamma_{f_0\to\gamma\gamma} &=& \frac{1}{32\pi \sqrt{p^2}} \frac{e^4}{8\pi^4 f_\pi^4 }
|B_K(p^2)\ \Psi(m_K^2; p^2; 0)
\nn\\
&&\quad + B_\pi(p^2)\ \Psi(m_\pi^2; p^2; 0)|^2,
 \enea

In deriving (\ref{width:agg}) and~(\ref{width:fgg}) the formula for the
width
 \bgea
\Gamma_{S\to\gamma\gamma} = 1/(2 \times 16\pi M_s) | \mathcal{M}_{S\to\gamma\gamma}
|^2
 \enea
is used, with the amplitude defined in~(\ref{eq:amplitudeA})
and~(\ref{eq:amplitudeF}), and symmetry factor $1/2$ for two
identical photons in the final state. Further, the sum over
polarization states $\lambda$ of the photon is performed  using
\begin{equation}
\sum_{\lambda= \pm 1} \epsilon^{(\lambda)}_\mu
\epsilon^{(\lambda)}_\nu{}^\ast  \to - g_{\mu\nu}
\end{equation}
under condition that polarization vector is contracted with the conserved
current.

The widths for $\phi(1020)$ meson decays are
 \bgea
\label{width:phigammaa0} \Gamma_{\phi \to \gamma a_0}&=&
\frac{1}{4\pi M_\phi}\frac{2}{3} \bigl( 1-
\frac{p^2}{M_\phi^2} \bigr) \Bigl[ \frac{\sqrt{2}eM_\phi
G_V}{f_\pi^4 (4\pi)^2} \Bigr]^2
\nn\\
&&\times\; \bigl| A_K(p^2)\ \Psi(m_K^2, p^2, M_\phi^2) \bigr|^2,
\\
\label{width:phigammaf0} \Gamma_{\phi \to \gamma f_0}&=&
\frac{1}{4\pi M_\phi}\frac{2}{3} \bigl( 1-
\frac{p^2}{M_\phi^2} \bigr) \Bigl[ \frac{\sqrt{2}eM_\phi
G_V}{f_\pi^4 (4\pi)^2} \Bigr]^2
\nn\\
&&\times\; \bigl|B_K(p^2)\ \Psi(m_K^2, p^2, M_\phi^2) \bigr|^2.
 \enea
Equations~(\ref{width:phigammaa0}) and~(\ref{width:phigammaf0})
are derived from general expression
 \bgea \label{eq:sec:5:1}
\Gamma_{\phi \to \gamma a_0/f_0}&=& \frac{\overline{ |\mathcal{M}_{\phi
\to \gamma S}|^2 }}{16\pi M_\phi} \bigl( 1- \frac{M_s^2}{M_\phi^2} \bigr)
,
 \enea
 with assumption (\ref{eq:FV-GV}).
The amplitude $\mathcal{M}_{\phi \to \gamma S}$ in given in (\ref{eq:rd2:2}).
The factor of $2/3$ in
(\ref{width:phigammaa0}) and (\ref{width:phigammaf0}) comes from the sum
over photon polarizations and average over vector-meson polarizations $\lambda$ by
means of
 \bgea
 \sum_{\lambda=0,\pm 1} E^{(\lambda)}_\mu E^{(\lambda)}_\nu{}^\ast
= - g_{\mu\nu} + \frac{Q_\mu Q_\nu}{M_\phi^2} .
 \enea

The widths for scalar mesons decay into photon and vector meson have the form
 \bgea
 \label{width:aToGammaV} \Gamma_{a_0 \to \gamma \rho/\omega}&=&
\frac{1}{2\pi \sqrt{p^2}} \bigl( 1-
\frac{M_{\rho/\omega}^2}{p^2} \bigr) \Bigl[ \frac{ e
M_{\rho/\omega} G_V}{f_\pi^4 (4\pi)^2} \Bigr]^2
\nn\\
&&\times\; \bigl| A_K(p^2)\ \Psi(m_K^2, p^2, M_{\rho/\omega}^2)
\bigr|^2,
\\
\label{width:fToGammaRho}
\Gamma_{f_0 \to \gamma \rho}&=&
\frac{1}{2\pi \sqrt{p^2}}
\bigl(
1- \frac{M_{\rho}^2}{p^2}
\bigr)
\Bigl[
\frac{ e M_{\rho} G_V}{f_\pi^4 (4\pi)^2}
\Bigr]^2
\nn\\
&&\times\; \bigl| B_K(p^2) \  \Psi(m_K^2, p^2, M_{\rho}^2)
\nn\\
&& + 2 B_\pi(p^2) \ \Psi(m_\pi^2, p^2, M_{\rho}^2) \bigr|^2,
\\
\label{width:fToGammaOmega} \Gamma_{f_0 \to \gamma \omega}&=&
\frac{1}{2\pi \sqrt{p^2}} \bigl( 1- \frac{M_{\omega}^2}{p^2}
\bigr) \Bigl[ \frac{ e M_{\omega} G_V}{f_\pi^4 (4\pi)^2} \Bigr]^2
\nn\\
&&\times\; \bigl| B_K(p^2)\ \Psi(m_K^2, p^2, M_{\omega}^2)
\bigr|^2,
 \enea
The expressions (\ref{width:aToGammaV}), (\ref{width:fToGammaRho}) and
(\ref{width:fToGammaOmega}) follow from
 \bgea
 \Gamma_{S\to \gamma V}
&=& \frac{1}{16 \pi M_s} \Bigl( 1 - \frac{M_V^2}{M_s^2} \Bigr)
\overline{ \Bigl| \mathcal{M}_{S\to \gamma V} \Bigr|^2 } ,
 \enea
where scalar meson mass is $M_s$, vector meson mass -- $M_V$, and
matrix element  $\mathcal{M}$ is given in~(\ref{M:aToGammaV}),
(\ref{M:fToGammaRho}) and (\ref{M:fToGammaOmega}) respectively.

Let us discuss difficulties one faces when trying to use
Eqs.~(\ref{width:agg})-(\ref{width:phigammaf0}) for fixing the couplings
$c_d, c_m, \tilde{c}_d, \tilde{c}_m$ and mixing angle $\theta$. It is
clear that accuracy and even existence of relevant experimental data are
very important. The particle properties are presented in
Table~\ref{table:particles} and the known decay widths in
Table~\ref{table:widths} (the latter contains also very recent data from
KEK~\cite{Mori_Belle}, however the errors are still too big).

\begin{table}
\caption{Decay data}
\label{table:widths}
\begin{center}
\begin{tabular}{llll}
\hline\noalign{\smallskip} $\frac{\Gamma_{\phi(1020)\to\gamma
f_0}}{\Gamma_{\phi,tot}}$ &=& $(4.40\pm0.21)\times 10^{-4}$
&\cite{pdg}\\
$\frac{\Gamma_{\phi(1020)\to\gamma a_0}}{\Gamma_{\phi,tot}}$ &=&
$(7.6\pm0.6)\times 10^{-5}$
&\cite{pdg}\\
$\frac{\Gamma_{\phi(1020)\to\gamma f_0}}{\Gamma_{\phi(1020)\to\gamma a_0}}$ &=& $6.1\pm0.6$
&\cite{pdg}\\
$\Gamma_{a_0\to\gamma\gamma}$ &=& $0.30 \pm 0.10 \;\text{keV}$
&\cite{pdg}\\
$\Gamma_{f_0\to\gamma\gamma}$ &=& $ 0.31 {}^{+ 0.08}_{- 0.11}\; \text{keV}$
&\cite{pdg}\\
$\frac{\Gamma_{a_0\to\eta\pi}
\Gamma_{a_0\to\gamma\gamma}}{\Gamma_{a_0,tot}}$ &=& $0.24 {}^{+
0.08}_{- 0.07}\;\text{keV}$
&\cite{pdg}\\
\hline\noalign{\smallskip}
$\Gamma_{f_0\to\pi\pi}$ &=& $34.2 {}^{+ 22.7}_{- 14.3}\; \text{MeV}$
&\cite{Mori_Belle}\\
$\Gamma_{f_0\to\gamma\gamma}$ &=& $0.205 {}^{+ 0.242}_{- 0.2}\; \text{keV}$
&\cite{Mori_Belle}\\
\noalign{\smallskip}\hline
\end{tabular}
\end{center}
\end{table}

At present there is a big ambiguity in the mass of
${\sigma={f_0}(600)}$ meson~\cite{pdg} (although one notices
smaller errors in reference from CLEO~\cite{MURAMATSU_02} in the
4th line in Table~\ref{table:particles}, given for overview
purpose only). For that reason we decided not to use information
on ${\sigma={f_0}(600)}$ neither in the coupling estimation nor in
the width prediction.

Strictly speaking~(\ref{width:akk}) and~(\ref{width:fkk}) are
valid if invariant mass of scalar meson $\sqrt{p^2}$ is bigger
than the pair production threshold $2 m_K$. In fact the physical
mass lies below threshold, i.e.  $2 m_K > M_{f_0}, \ M_{a_0}$
(Table~\ref{table:particles}). This makes impossible to use
(\ref{width:akk}) and~(\ref{width:fkk}) directly in the fitting
procedure (in principle, one can use an approximate
method~\cite{Maiani:2004uc}).

From Table~\ref{table:widths} one sees that the precision of
estimate for $\Gamma_{a_0\to\pi\eta}$ depends on accuracy with
which the total $a_0$ width $\Gamma_{a_0,tot}$ is known, as only
the ratio
 \bgea
 \label{eq:pietaproblem} \frac{\Gamma_{a_0\to\eta\pi}
\Gamma_{a_0\to\gamma\gamma}}{\Gamma_{a_0,tot}}
 \enea
is measured~\cite{pdg}. Unfortunately $\Gamma_{a_0,tot}$ has big
experimental error and therefore extraction of
$\Gamma_{a_0\to\pi\eta}$ from (\ref{eq:pietaproblem}) leads to a
big error. Thus this information should not be used in the
analysis.

Further, formally (\ref{width:agg}) and~(\ref{width:phigammaa0})
are not independent as they are expressed through the same factor
$A_K$. We prefer to use (\ref{width:agg}) because of a non-trivial
assumption (\ref{eq:FV-GV}) for the couplings used in derivation
of (\ref{width:phigammaa0}). For realistic values of $F_V$ and $G_V$
the relation (\ref{eq:FV-GV})
is satisfied only approximately (see Appendix~\ref{App:A}).

From the above reasonings it becomes clear that fixing five parameters in
question is not an easy task. Therefore we reduce the number of
independent parameters from five to three by applying the large-$N_c$
relations~(\ref{eq:largeNC}). These relations are briefly discussed in
Appendix~\ref{App:B}. Then to find values of $c_d, c_m, \theta$ one can
use (\ref{width:agg}),~(\ref{width:fgg}) and~(\ref{width:phigammaf0}).

In the analysis below we take masses of $a_0(980)$ and $f_0(980)$ equal
and put $\sqrt{p^2}\approx\;980\ \text{MeV}$.
Let $A_K$, $B_K$ and $B_\pi$ be our estimates for $A_K(M_{a_0}^2)$, $B_K(M_{f_0}^2)$ and $B_\pi(M_{f_0}^2)$.
Applying the
constraint~(\ref{eq:largeNC}) to (\ref{eq:four_coef}) we find that the
scalar mixing angle~$\theta$ satisfies the equation
 \bgea
\label{eq:eqaution_theta}
 4\mu \; \cos \theta + \sqrt{2}\; \sin\theta &=& \sqrt{6}\; \frac{B_K}{A_K}
,
 \enea
where $\mu = \pm 1$ stands for two possible choices of sign
in~(\ref{eq:largeNC}). For the coupling constants
 \bgea
\label{eq:chir_coupl_form} c_d &=& \frac{\sqrt{2}}{p^2 (m_K^2 - m_\pi^2)}
\bigl( m_\pi^2 A_K - m_K^2 R B_\pi \bigr)
,\nn\\
c_m &=&
\frac{-1}{\sqrt{2} p^2 (m_K^2 - m_\pi^2)}
\bigl(
(p^2 - 2m_\pi^2) A_K
\nn\\
&& - (p^2 - 2m_K^2) R B_\pi
\bigr)
,\enea
where
\bgea
R^{-1} &\equiv& \frac{B_K}{A_K}
 - \sqrt{3} \;\sin \theta .
 \enea

The values of $A_K$, $B_K$ and $B_\pi$ in~(\ref{eq:four_coef}),
extracted
from~(\ref{width:agg}),~(\ref{width:fgg}),~(\ref{width:phigammaf0})
and experiment, are
 \bgea
\label{eq:four_coef_estimate}
B_K   &\approx& 3.4716 \times 10^7 \ \text{MeV}^3 ,\nn\\
B_\pi &\approx& (1.029 \; \text{or}\; 4.96)\times 10^7 \ \text{MeV}^3,\nn\\
A_K   &\approx& 2.2456 \times 10^7 \ \text{MeV}^3 ,
 \enea
(here $\text{Im}\, B_\pi \equiv 0$ is assumed). Inserting these
values in~(\ref{eq:eqaution_theta}) and~(\ref{eq:chir_coupl_form})
one obtains couplings and mixing angle in
Table~\ref{table:coup_estim_1}. The relation between the mixing
angle~$\theta$ for $\mu = +1$ and~$\mu = -1$ is discussed in
Appendix~\ref{App:B}.

\begin{table*}
\caption{Chiral couplings and mixing angle}
\label{table:coup_estim_1}
\begin{center}
\begin{tabular}{r|rccc}
\hline\noalign{\smallskip}
$B_\pi,\; (10^7 \ \text{MeV}^3)$ & $1.029 $ & $1.029 $ & $4.96$ & $4.96$
\\
\noalign{\smallskip}\hline\noalign{\smallskip}
$c_d\ (\text{MeV})$    &$-6.39$&$-52.57$&$-41.67$&$-264.37$
\\
$c_m\ (\text{MeV})$    &$-58.83$&$-13.15$&$-23.93$&$196.37$
\\
$\theta$ \, (\text{for} $\mu = +1)$ & $-7.329^\circ$ & $46.271^\circ$ &
$-7.329^\circ$ & $46.271^\circ$
\\
$\theta$ \, (\text{for} $\mu = -1$)& $-172.671^\circ$ & $133.729^\circ$ &
$-172.671^\circ$ & $133.729^\circ$
\\
\noalign{\smallskip}\hline
\end{tabular}
\end{center}
\end{table*}

\subsection{Analysis of loop integrals}
\label{sec:loop-anal}

\begin{figure*}
\begin{center}
  \begin{tabular}{ccc}
    \resizebox{0.3\textwidth}{!}{
    \includegraphics{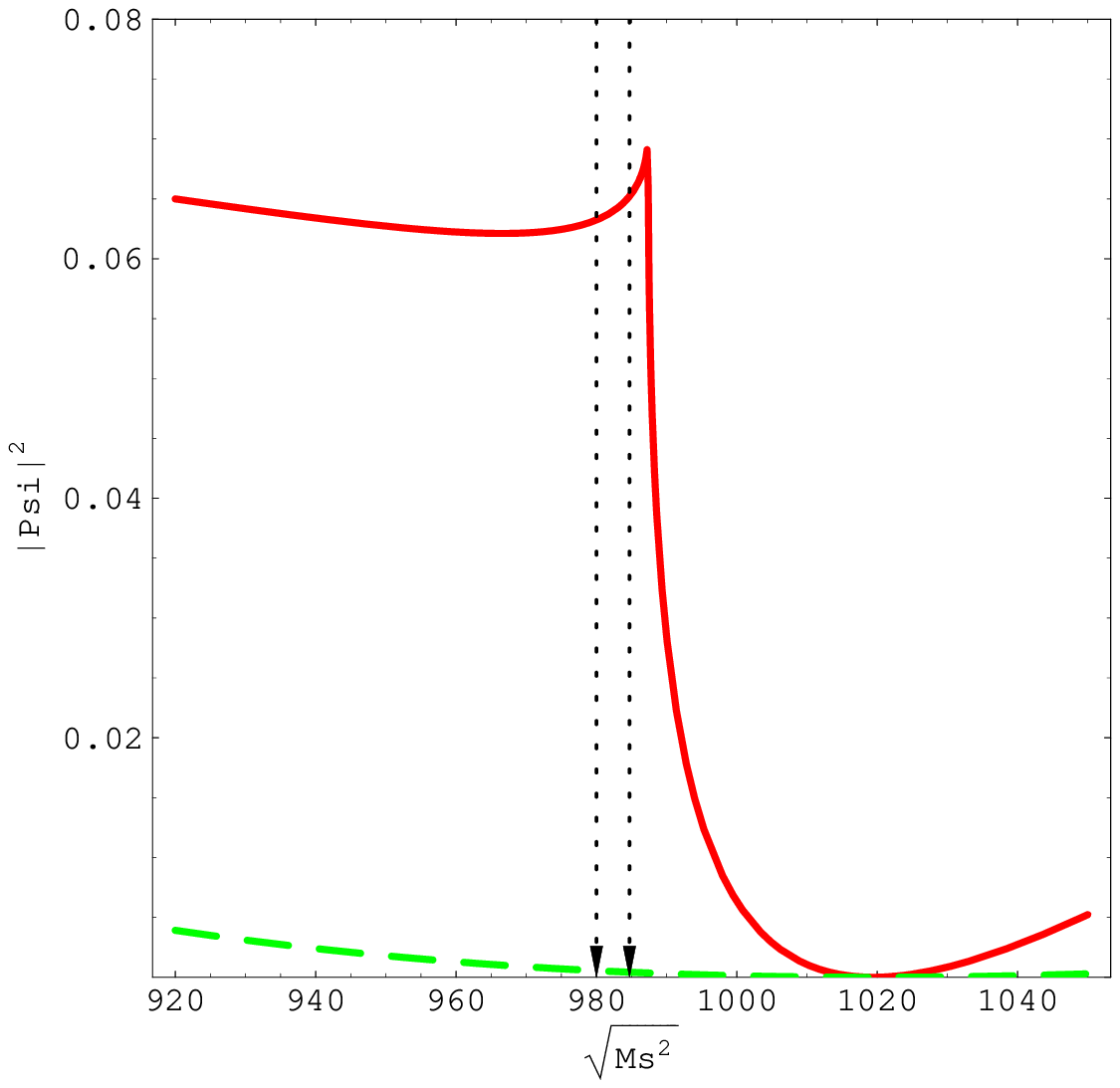}
    }
  &
    \resizebox{0.3\textwidth}{!}{
    \includegraphics{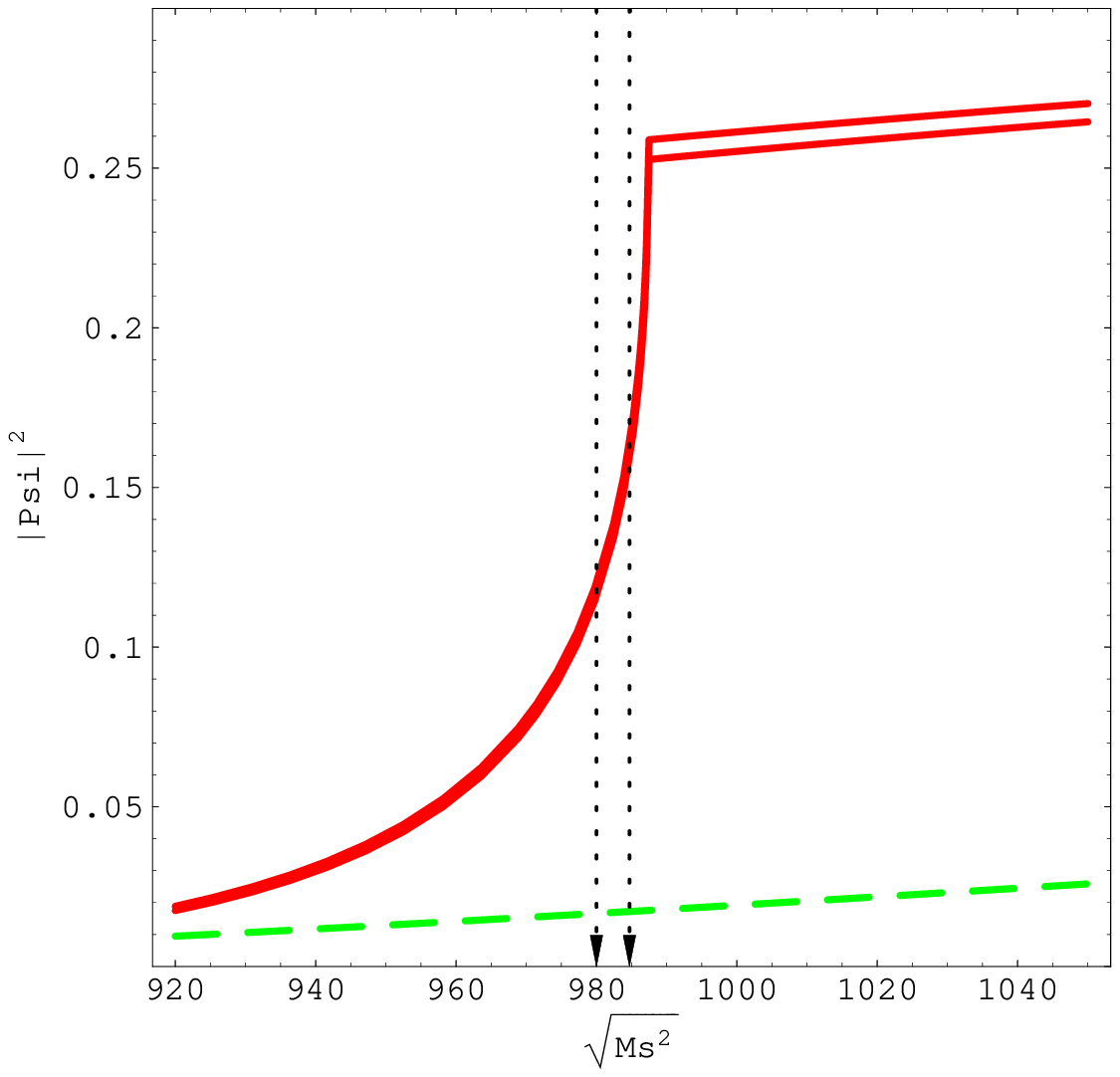}
    }
  &
    \resizebox{0.3\textwidth}{!}{
    \includegraphics{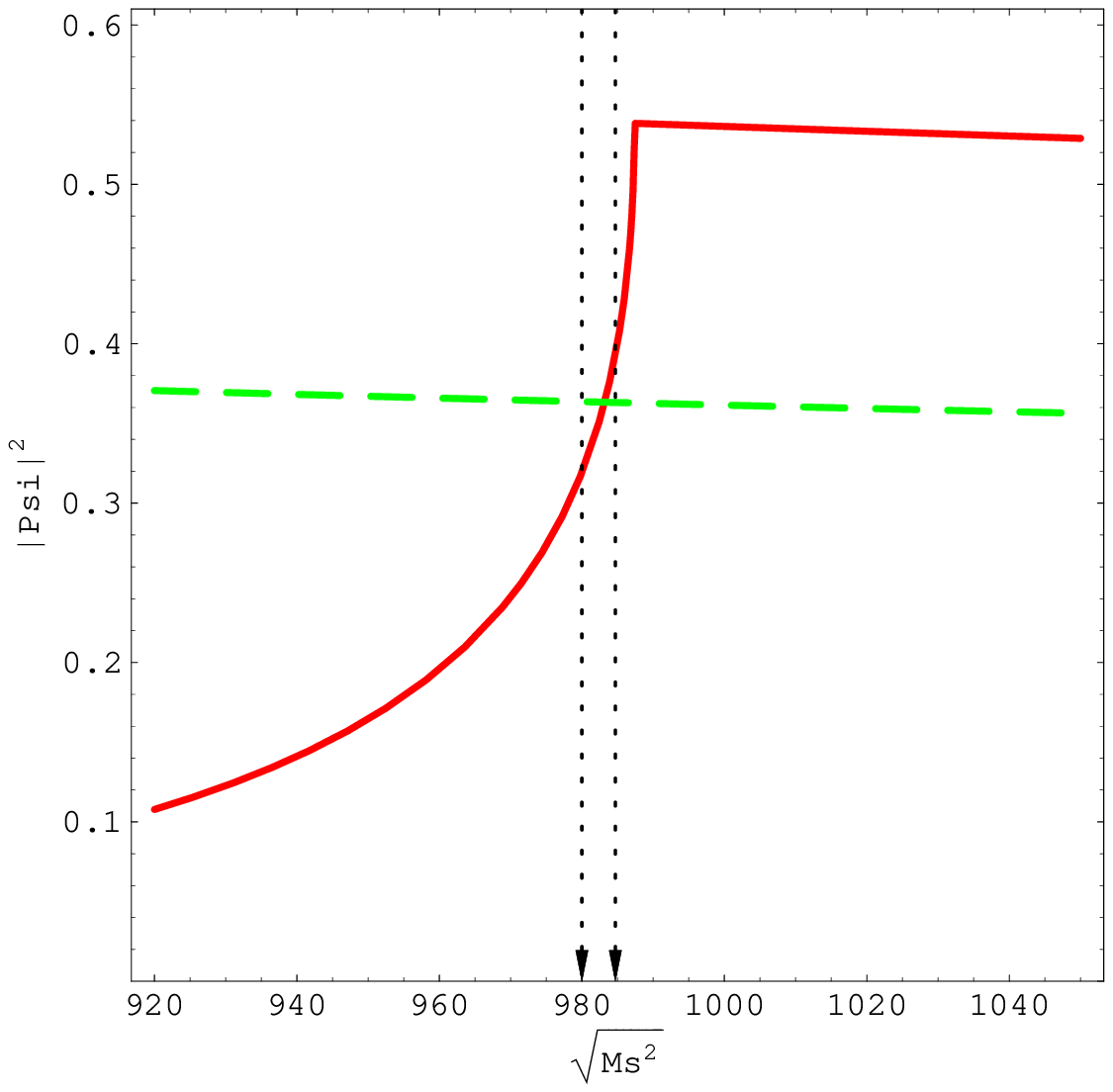}
    }
  \\
  (a)~$\begin{matrix}
  \phi \to (K\bar{K} \to) \ \gamma \ a_0 \\
  \phi \to (K\bar{K} \to) \ \gamma \ f_0 \\
  \phi \to (\pi^+\pi^- \to) \ \gamma \ f_0
  \end{matrix}$
  &
  (b)~$\begin{matrix}
  a_0 \ \to (K\bar{K} \to) \ \gamma \ \rho \\
  a_0 \ \to (K\bar{K} \to) \ \gamma \ \omega \\
  f_0 \ \to (K\bar{K} \to) \ \gamma \ \rho \\
  f_0 \ \to (\pi^+\pi^- \to) \ \gamma \ \rho \\
  f_0 \ \to (K\bar{K} \to) \ \gamma \ \omega
  \end{matrix}$
  &
  (c)~$\begin{matrix}
  a_0 \ \to (K\bar{K} \to) \ \gamma \gamma \\
  f_0 \ \to (K\bar{K} \to) \ \gamma \gamma \\
  f_0 \ \to (\pi^+\pi^- \to) \ \gamma \gamma
  \end{matrix}$
  \end{tabular}
\end{center}
\caption{Comparison of loop integrals squared, $|\Psi|^2$. Solid line is
drawn for kaon loop, dashed line for pion loop. Vertical dotted lines
mark \textbf{assumed} physical values of scalar meson mass ($M_{f_0}$ and $M_{a_0}$). Two
solid lines in (b) account for different masses of $\rho$ and $\omega$
mesons. Interference between kaon and pion loops is not included. }
\label{fig:integral_contr_1}
\end{figure*}

Here we analyze dependence of the loop
integrals~(\ref{eq:psi_defin1}),~(\ref{eq:psidef2}),~(\ref{eq:integrals_close_jue1})
and~(\ref{eq:integrals_close_jue2}) on the scalar meson invariant
mass. Firstly, such dependencies are important for processes,
where the off-shell scalar resonances enter, i.e. for any scalar
meson production and its subsequent decay to $\gamma\gamma$ or
$\gamma \ \rho/\omega$. Secondly, the masses of $a_0$ and $f_0$
mesons are not accurately established yet. New experimental
results may alter considerably the existing values and it is
important to know how results of an approach depend on the masses.

Notice that our definition of loop integral $\Psi$ automatically
includes the loop kinematic factors, i.e. $\Psi = (a-b) I(a,b)$ in
terms of analytic approach~\cite{Close93}. The quantity $|\Psi|^2$ is
convenient in analysis of loop contribution to decay
probabilities~(\ref{eq:amplitudeA}), (\ref{eq:amplitudeF})
(\ref{eq:rd2:2}) and~(\ref{M:aToGammaV})-(\ref{M:fToGammaOmega}).

Fig.~\ref{fig:integral_contr_1} shows dependence of $|\Psi|^2$ on
the scalar-meson invariant mass. This figure does not include any
possible interference effects between the pion and kaon loops.
From Fig.~\ref{fig:integral_contr_1}~(c) one sees that the pion
contribution to $f_0 \ \to \gamma \gamma$ decay turns out as large
as the kaon one. The loop integrals crucially depend on the
pseudoscalar threshold and the relation between the masses of
pseudoscalar and vector particles. Especially dependencies for the
kaon loops are complex due to proximity of the $K\bar{K}$
threshold. Two dotted vertical lines in
Fig.~\ref{fig:integral_contr_1} show the physical masses of scalar
mesons. It is seen that the kaon-loop contribution rapidly changes
near the $K\bar{K}$ threshold in the vicinity of $M_{a_0}$
($M_{f_0}$), therefore an error in the mass value may cause
drastic changes in the $K\bar{K}$ contribution.

Note that earlier similar loop integrals for $\phi \to \gamma \ a_0/f_0$
and $a_0/f_0 \ \to \gamma \ \rho/\omega$ decays were analyzed in
~\cite{Achasov_Ivanchenko,Hanhart,Close93}. In
particular, authors of~\cite{Hanhart} concluded that the pion loops gave
negligible contribution to the decays.

\begin{figure} \begin{center}
\resizebox{0.5\textwidth}{!}{
 \includegraphics{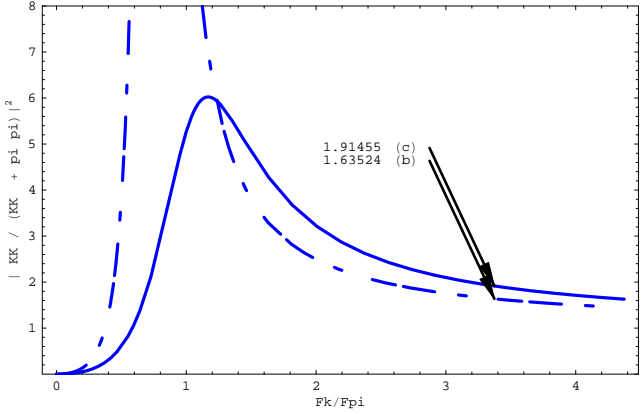}
}
\end{center}
\caption{ Relative kaon loop contribution, $\left| (K\bar{K}) /
(\pi^+\pi^- \ + \ K\bar{K}) \right|^2$, to (\ref{eq:amplitudeF})
and (\ref{M:fToGammaRho}) versus ratio $B_K / B_\pi$. Curves (b)
and (c) correspond to decays $f_0 \to\gamma \rho $ and $f_0 \to
\gamma\gamma$ respectively (see also legend in
Fig.~\ref{fig:integral_contr_1}). Our fit gives $B_K / B_\pi =
3.37$. } \label{fig:integral_contr_2}
\end{figure}

In this connection we stress that for any observable not only the
loop integrals but also the couplings matter. Thus it is important
to compare the pion and kaon contributions taking into account the
corresponding coupling constants as well as the interference
effects. Fig.~\ref{fig:integral_contr_2} shows the ratio of the
$K\bar{K}$ contribution to the total $K\bar{K} + \pi \pi$
contribution calculated from (\ref{eq:amplitudeF}),
(\ref{M:fToGammaRho}) and Table~\ref{table:psi}. This ratio
exhibits the effect of interference between pion and kaon loops
and depends on the ratio $B_K / B_\pi$. Our estimates for the
couplings lead to $B_K / B_\pi = 3.37$, and arrows in
Fig.~\ref{fig:integral_contr_2} mark the values of relative kaon
contribution
 \bgea
\left| \frac{B_K\ \Psi(m_K^2, M_f^2, M_\rho^2) }{2\ B_\pi \ \Psi(m_\pi^2,
M_f^2, M_\rho^2) + B_K\ \Psi(m_K^2, M_f^2, M_\rho^2)} \right|^2 &=&1.635
,
\nn\\
\left| \frac{B_K\ \Psi(m_K^2, M_f^2, 0) }{B_\pi \ \Psi(m_\pi^2, M_f^2, 0)
+ B_K\ \Psi(m_K^2, M_f^2, 0)} \right|^2 &=&1.915 \ \ \ \nn
 \enea
to decays $f_0 \to\gamma \rho $ and $f_0 \to \gamma\gamma$ respectively.

 Results in Fig.~\ref{fig:integral_contr_1}~(a)
for the $\phi$ decays strongly favor the kaon loops compared to
the pion loops. That would be an argument, additional to OZI
suppression rule, for not using pion loops for $\phi$ decays. For
other processes Fig.~\ref{fig:integral_contr_2} gives an adequate
measure of the pion-kaon concurrence. In $f_0 \ \to \ \gamma \
\rho $ decay omitting the pion loops would lead to $\approx 60\%$
overestimate of the width. The pion loops are very important in
the two-photon decay of $f_0$: they reduce the decay rate by a
factor of $1.9$.


\subsection{Model predictions}

In the present model two types of predictions are obtained. On the one hand,
the observables depend on values of the model parameters, and thus can be
evaluated after specific values are chosen.

On the other hand, several ratios of the widths turn out to be
independent of specific values of couplings $c_m$, $c_d$,
$\tilde{c}_m$, $\tilde{c}_d$ and $\theta$. We find three such
ratios which are constant in the present model for any values of
these parameters:
 \bgea
  \label{eq:PredNoDep1}
\frac{\Gamma_{a_0\to\gamma\gamma}}{\Gamma_{\phi\to \gamma a_0}} &=& 0.422
\\
&=& \frac{ 3 e^2 f_\pi^4  M_{\phi}}{G_V^2 M_{a_0}(M_{\phi}^2 -
M_{a_0}^2)} \left| \frac{\Psi(m_K^2; M_{a_0}^2; 0)}{\Psi(m_K^2;
M_{a_0}^2; M_\phi^2)} \right|^2 \nn ,
 \enea
From experiment (Table~\ref{table:widths}) one obtains about
$0.93$ for this ratio.

Another ratio
 \bgea
  \label{eq:PredNoDep2}
\frac{\Gamma_{a_0\to\gamma\rho}}{\Gamma_{a_0\to\gamma\omega}} &=& 1.043
\\
&=& \frac{(M_{a_0}^2 - M_\rho^2)M_\rho^2}{(M_{a_0}^2 - M_\omega^2) M_\omega^2}
\left| \frac{ \Psi(m_K^2; M_{a_0}^2;
M_\rho^2) }{  \Psi(m_K^2; M_{a_0}^2; M_\omega^2) } \right|^2 \nn
 \enea
has not been measured so far, though theoretical predictions
exist. In particular, it was shown~\cite{Hanhart} that the
quark-loop mechanism in the two-quark model gives value about
$1/9$, the four-quark structure leads to $\approx 0$, while
kaon-loop mechanism produces almost equal $a_0\to\gamma \ \rho$
and $a_0\to\gamma \ \omega$ widths. Our result appears close to
the latter prediction.

From (35), (39) and (43) it is also possible to derive
 \bgea
\frac{\Gamma_{a_0\, \to\, \gamma\, \rho (\omega)}}{\Gamma_{\phi\,
\to\, \gamma\, a_0}} &\approx& 12
\\\nonumber
&=&
\frac{3 M_\phi^3}{M_{a_0}^3}
\frac{M_{a_0}^2 - M_{\rho (\omega)}^2}{M_{\phi}^2 - M_{a_0}^2}
\left( \frac{M_{\rho (\omega)}^2}{2 M_\phi^2} \right)
\\
\nonumber & \times & \, \left| \frac{\Psi(m_K^2, M_{a_0}^2,
M_{\rho (\omega)}^2)}{\Psi(m_K^2, M_{a_0}^2, M_\phi^2)} \right|^2
. \enea Analogous ratio can be deduced from results of~[16]. It
appears to be approximately $5.6$. The difference may be addressed
to dissimilarity between the models, particular choice of mass and
coupling constant values.

Now, let us focus on the coupling-dependent results calculated
according to (\ref{width:agg})-(\ref{width:fToGammaOmega}),
(\ref{width:phigammaa0}) and (\ref{width:phigammaf0}).
Tables~\ref{table:width_estim_STRONG} and~\ref{table:width_estim}
show predictions of the model and comparison with available data.
Table~\ref{table:width_estim_STRONG} presents obtained values of
$c_d$, $c_m$ and $\theta$, and strong decay widths for $a_0$ and
$f_0$ mesons.

In Table~\ref{table:width_estim} one can see results for radiative
decay widths.
The column \textbf{I} shows calculations with our parameters $c_d
= -6.38\ \text{MeV}$, $c_m = -58.83\ \text{MeV}$ and $\theta =
-7.33^\circ$. The entries in Table~\ref{table:width_estim}, which
were taken as input in the fitting procedure, are marked with
asterisk ($\Gamma_{\phi \to \gamma f_0}/ \Gamma_\phi$,
$\Gamma_{a_0 \to \gamma \gamma}$ and $\Gamma_{f_0 \to \gamma
\gamma}$). The column \textbf{II} is calculated with an ``ideal''
mixing angle $\theta = -35.3^\circ$ \ ($\sin \theta = -
1/\sqrt{3}$, $\cos \theta = \sqrt{2/3}$) as chosen
in~\cite{Fajfer}. In this case decay $\sigma = f_0(600) \to
\pi\pi$ is forbidden\footnote{In $q \bar{q}$ quark model this case
corresponds to $f_0(600) = s \bar{s}$}, though in fact it should
be super-allowed. The column \textbf{III} deals with another
``ideal'' mixing angle $\theta = 54.7^\circ$ \ ($\cos \theta =
1/\sqrt{3}$, $\sin \theta = \sqrt{2/3}$). In this case decay
$f_0(980) \to \pi\pi$ turns out forbidden in contradiction with
experimental evidence. Therefore the choices~\textbf{II} and~\textbf{III}
do not look realistic.

\begin{table*}
\caption{Strong decays of scalar mesons}
\label{table:width_estim_STRONG}
\begin{center}
\begin{tabular}{r|lll|l}
\hline\noalign{\smallskip}  & Our estimate (\textbf{I}) &
Estimate (\textbf{II}) & Estimate (\textbf{III}) &Exp. value~\cite{pdg}
\\
\noalign{\smallskip}\hline\noalign{\smallskip} $c_d ,
\;\text{MeV}$ & $-6.38$ & $\pm32$ & $\pm32$ & -
\\
$c_m , \;\text{MeV}$ & $-58.83$ & $\pm42$ & $\pm42$ & -
\\
$\theta$ & $-7.33^\circ$ & $ -35.26^\circ$ &  $ 54.73^\circ$ & -
\\
\hline\noalign{\smallskip} $\Gamma_{a_0 \to \pi\eta} ,
\;\text{MeV}$ & $14.2$ & $172.4$ & $172.4$ & -
\\
$\Gamma_{f_0 \to \pi\pi}, \;\text{MeV}$ & $41.8$ & $775.7$ & $7.4
\times 10^{-6}$ & $34.2 {}^{+ 22.7}_{- 14.3}$
\\
$\Gamma_{a_0,tot}, \;\text{MeV}$ & $17.8 $ & $215.4$ & $215.4$ &
$50-100$
\\
\noalign{\smallskip}\hline
\end{tabular}
\end{center}
\end{table*}

\begin{table*}
\caption{Model predictions and available data for electromagnetic decays.
(Asterisk ${}^*$ marks the experimental values, used for extraction of couplings.
For couplings used in columns (\textbf{I}), (\textbf{II}) and (\textbf{III})
see Table~\ref{table:width_estim_STRONG})
}
\label{table:width_estim}
\begin{center}
\begin{tabular}{l|llllll|l}
\hline\noalign{\smallskip} Observable & Our est. (\textbf{I}) &
Est. (\textbf{II}) & Est. (\textbf{III}) & \cite{Hanhart}
(\textbf{IV}) & \cite{Black} (\textbf{Va}) & \cite{Black} (\textbf{Vb}) & Exp. value~\cite{pdg}
\\
\noalign{\smallskip}\hline\noalign{\smallskip} $\frac{\Gamma_{\phi
\to \gamma a_0}}{\Gamma_{\phi,tot}},\ 10^{-4} $ & $1.67 $ & $2.13
$ & $2.13$ & $1.4 $ & $-$ & $-$ & $(7.6 \pm 0.6) \times 10^{-1} $
\\
$\frac{\Gamma_{\phi \to \gamma f_0}}{\Gamma_{\phi,tot}},\ 10^{-4}
$ & $4.40
$$^*$ & $2.31$ & $4.63$
& $1.4$ & $4.92\pm 0.07$ & $4.92\pm 0.07$ & $4.40 \pm 0.21$
\\
$\frac{\Gamma_{\phi \to \gamma f_0}}{\Gamma_{\phi \to \gamma
a_0}}$ & $2.64$ & $1.08$ & $2.17$ & $1$ & $0.26 \pm 0.06$ & $0.46
\pm 0.09$ & $6.1 \pm 0.6$
\\
\noalign{\smallskip}\hline\noalign{\smallskip} $\Gamma_{a_0 \to
\gamma\gamma},\; \text{keV}$ & $0.30$$^*$ & $0.383 $ & $0.383 $ &
$0.24$ & $0.28\pm 0.09$$^*$ & $0.28\pm 0.09$$^*$ & $0.30 \pm 10$
\\
$\Gamma_{f_0 \to \gamma\gamma}, \;\text{keV}$ & $0.31 $$^*$ &
$0.323$ & $0.62$ & $0.24$ & $0.39\pm 0.13$$^*$ & $0.39\pm
0.13$$^*$ & $0.31_{-0.07}^{+0.08}$
\\
\noalign{\smallskip}\hline\noalign{\smallskip} $\Gamma_{a_0 \to
\gamma \rho}, \; \text{keV}$ & $9.1 $ & $11.65$ & $11.65$ & $3.4 $
& $3.0 \pm 1.0$ & $3.0 \pm 1.0$ & $-$
\\
$\Gamma_{f_0 \to \gamma \rho},\; \text{keV}$ & $9.6 $ & $0.95$ &
$16.6 $ & $3.4 $ & $19 \pm 5$ & $3.3\pm 2.0$ & $-$
\\
$\Gamma_{a_0 \to \gamma \omega},\; \text{keV}$ & $8.7 $ & $11.15$
& $11.15$ & $3.4$ & $641 \pm 87$ & $641 \pm 87$ & $-$
\\
$\Gamma_{f_0 \to \gamma \omega}, \; \text{keV}$ & $15.0 $ & $7.93
$ & $15.85$ & $3.4$ & $126 \pm 20$ & $88\pm 17$ & $-$
\\
\noalign{\smallskip}\hline
\end{tabular}
\end{center}
\end{table*}

For comparison of our results with predictions of other models we
add columns \textbf{IV}, \textbf{Va} and \textbf{Vb}. In
particular, kaon-loop model~\cite{Hanhart} (\textit{KLM}) is
selected (column \textbf{IV}), which is somewhat similar to the
present calculation. In columns \textbf{Va} and \textbf{Vb}
predictions of vector-meson-dominance (\textit{VMD})
model~\cite{Black} (Table~I therein) are shown for two different
sets of parameters. The authors apply a chiral Lagrangian with
strong trilinear scalar-vector-vector interaction.

\subsection{Discussion}
In this subsection we briefly compare our results with those of
\textit{KLM}~\cite{Hanhart} and \textit{VMD}~\cite{Black} models
and comment on correspondence of predicted widths to experiment.

As it is seen from Table~\ref{table:width_estim}, our model,
contrary to \textit{KLM}, gives not only the rate for decay of a
given type ($S\to\gamma\gamma$, $S\to \gamma V$ and $\phi\to
\gamma S$ groups) but also different decay rates for $a_0$ and
$f_0$ mesons. For some of the channels the results of \textit{KLM}
are qualitative estimates, corrections~\cite{Hanhart:KKMol-gg} to
which should be calculated as discussed in~\cite{Hanhart}.
Nevertheless, our results for $S\to\gamma\gamma$ and $\phi\to
\gamma S$ are in agreement with \textit{KLM} within an order of
magnitude. For the ratio ${\Gamma_{a_0\to\gamma\gamma}} /
{\Gamma_{\phi\to \gamma a_0}}$ we also get a close value, which
approximately corresponds to experimental result. At the same time
we obtain the widths for $S\to\gamma V$ decays which are bigger
than the values in the \textit{KLM}. The latter discrepancy is due
to $SU(3)$ relations for strong interaction (see (\ref{eq:F4}))
and our couplings of $\phi$ and $\rho/\omega$ mesons to $K\bar{K}$
turn out to be different from those used in~\cite{Hanhart}.

Regarding the \textit{VMD} model~\cite{Black}, one can see from
Table~\ref{table:width_estim} that quite a big decay widths for $S
\to \gamma \omega$ are obtained there compared to our results,
while $S \to \gamma \rho$ predictions differ not so much. Note
also a big difference in the values of the ratio ${\Gamma_{\phi
\to \gamma f_0}}/{\Gamma_{\phi \to \gamma a_0}}$.

From results presented in Tables~\ref{table:width_estim_STRONG}
and~\ref{table:width_estim} one concludes that predictions for
scalar meson decay widths are very sensitive to model details.
Therefore the future experiments in which these processes will be
studied may help to discriminate between different models of
scalar mesons.

In general, for comparison with experiment, in which scalar
resonances contribute, a more appropriate observable is the
invariant mass distribution. As an example consider the reaction
$e^+ e^- \to \gamma^* \to \pi \pi \gamma$ at the CM energy close
to the $\phi (1020)$ mass~\cite{KLOEresults}. This reaction allows
for extraction of the branching ratio
\begin{equation}
\frac{\mathrm{d} B_{\phi \to \pi \pi \gamma}}{\mathrm{d} p^2}
=\frac{1}{\Gamma_{\phi, \, tot}} \frac{\mathrm{d}
\Gamma_{\phi \to \pi \pi \gamma} }{\mathrm{d} p^2},
\label{eq:BR_1}
\end{equation}
where $p^2$ is the two-pion invariant mass squared.
Within the present framework this branching ratio can be
calculated from
\begin{equation}
\frac{\mathrm{d} \Gamma_{\phi \to \pi \pi \gamma}}{\mathrm{d} p^2}
=\Gamma_{\phi \to \gamma f_0 }(p^2) B_{f_0 \to \pi\pi} (p^2) \times
\big(-\frac{1}{\pi}\big) \mathrm{Im} D_{f_0} (p^2). \label{eq:BR_2}
\end{equation}
Here $\Gamma_{\phi \to \gamma f_0 } (p^2)$ is the
$ \phi \to \gamma f_0 $ decay width (\ref{width:phigammaf0})
for arbitrary $p^2$,  \
$D_{f_0} (p^2) = [p^2 - m_f^2 + i m_f \Gamma_{f_0, \text{tot}}
(p^2)]^{-1}$ is the scalar-meson propagator and branching ratio $B_{f_0
\to \pi\pi} (p^2) = \Gamma_{f_0 \to \pi \pi} (p^2) / \Gamma_{f_0,
\text{tot}}(p^2)$ relates $f_0 \to \pi \pi$  decay width
(\ref{width:fpp}) to the total $f_0$ width $\Gamma_{f_0,
\text{tot}}(p^2)$. A more advanced form of the propagator including both
real and imaginary parts of the self-energy was suggested
recently~\cite{Achasov:2004uq}.
The problem of finite resonance width effects in invariant mass distributions for
$\pi^0\pi^0$ and $\pi^0\eta$ in the $\phi$ radiative decays
is important~\cite{Oller:2002na}.

Note also that in Ref.~\cite{Hanhart} (Appendix) a more general
distribution over invariant masses of both the initial and final
resonances for $S \to \gamma V$ decays is discussed.

The scalar octet and singlet mixing angle $\theta$ appeared to be
crucial parameter in the fit. We should remark that a detailed
study of mixing angle was performed in~\cite{Oller:2003vf} using
the inverse amplitude method. The basic processes there were
elastic $\pi\pi$, $\pi\eta$, $K\bar{K}$ and $K\eta$ scattering and
angle value was different from our estimate\footnote{The
definition of mixing angle in~\cite{Oller:2003vf} is also
different.}. There are also models in which $f_0$ is mainly the
singlet state with $\theta \approx 0$, for example, an application
of the Bethe-Salpeter equation with a linear confinement $q
\bar{q}$ potential to calculation of scalar-meson mass
spectrum~\cite{Klempt_spectrum}.

This discrepancies, to our opinion, may not be caused only by
differences in the applied models. The problem may be related to a
non-trivial structure and behavior of light scalar resonances. As
it was emphasized by Bugg~\cite{Bugg:2004xu}, unification of
observed resonances in elastic scattering experiments with
corresponding ones seen in radiative decays is a big challenge.
They show up in different ways, and it would be important to build
a consistent bridge between these and those properties of
resonances.

\subsection{Possible interactions beyond the model}
Comparison of our results with predictions of other models,
especially the results independent of the choice of couplings,
shows that the present model does not allow one to reproduce the
ratio ${\Gamma_{a_0\to\gamma\rho}} /{\Gamma_{a_0\to\gamma\omega}}
=1/9$, which is obtained in $q \bar{q}$ model and quark-loop
mechanism~\cite{Hanhart}, or ${\Gamma_{a_0\to\gamma\rho}}
/{\Gamma_{a_0\to\gamma\omega}} \approx 0$ in $q q \bar{q} \bar{q}$
model~\cite{Hanhart}. The present model is insensitive to the
structure of scalars.

At this point one can think of a direct (or contact) coupling of scalars
to two photons as an extension of the present model. Similar terms were
introduced in~\cite{Fajfer} and have the order $\mathcal{O}(p^4)$
\begin{equation}
  L_{1} = g \left\langle S^{oct} f_+^{\mu\nu} f_{+\mu\nu}  \right\rangle
  +   g^{\prime} S^{sing} \left\langle f_+^{\mu\nu} f_{+\mu\nu}  \right\rangle,
\label{new_S_gamma_gamma}
\end{equation}
where $g, \ g^{\prime}$ are coupling constants and $f_+^{\mu\nu}$ is
defined in (\ref{eq:f+}).

Analogously, constructing the $C$ and $P$ invariant terms with
$V_{\mu\nu}$ one can propose the $S \gamma V $ interactions
\begin{equation}
  L_{2} = g^{\prime \prime} \left\langle S^{oct} f_{+}^{\mu\nu}  V_{\mu\nu} \right\rangle
  + g^{\prime \prime \prime} S^{sing}\left\langle  f_{+}^{\mu\nu} V_{\mu\nu}
  \right\rangle
  \label{new_S_gamma_V}
\end{equation}
which are bilinear in resonance fields. Lagrangian (\ref{new_S_gamma_V})
has the order $\mathcal{O}(p^2)$ and contains two more couplings
$g^{\prime \prime}, \ g^{\prime \prime \prime}$. Of course, these terms
do not violate chiral symmetry. Four additional coupling constants $g,
g^\prime, g^{\prime \prime}, g^{\prime \prime \prime}$ should be fixed
from certain observables.

To our opinion, Lagrangians (\ref{new_S_gamma_gamma}) and
(\ref{new_S_gamma_V}) can be useful in phenomenological descriptions of
scalar radiative decays. They may represent effects related to specific
quark structure of the scalar mesons, which is not accounted for in
chiral Lagrangian $L^B$ (\ref{lagr:general}). This aspect lies beyond the
scope of the present paper.


\section{Conclusions}
\label{sec:conclusions}

Within ChPT with vector and scalar mesons~\cite{EckerNP321} we have
calculated the radiative decays $a_0\to\gamma\gamma$, $f_0
\to\gamma\gamma$, $\phi \to \gamma a_0$ and $\phi \to \gamma f_0$.
These decays and corresponding invariant mass distributions
can be measured in $e^+e^-$ annihilation in Frascati by
KLOE~\cite{Ambrosino:2006gk,Ambrosino:2005wk} and
Novosibirsk with VEPP-2000.

The derivative and non-derivative couplings of scalar mesons to
pseudoscalar ones are consistently included. The gauge invariance
of the amplitudes and cancelation of divergencies from different
diagrams are explicitly demonstrated. The obtained amplitudes are
finite without counter terms. For $\phi$ decays, in addition, we
used the relation $F_V = 2\; G_V$ between electromagnetic and
strong couplings of vector mesons in order to get rid of
divergencies. This relation was previously discussed in
\cite{EckerPLB223} in connection with alternative approaches,
Hidden Local Gauge Symmetry~\cite{HGS} and massive Yang-Mills
models~\cite{MYM}. Note, that this relation does not follow from
chiral symmetry but does not contradict it as
well~\cite{EckerPLB223}.

The scalar flavor singlet-octet mixing angle $\theta$ is obtained
from the fit, as well as estimates for octet chiral couplings
$c_m$ and $c_d$. It should be noted that the values of these
parameters strongly correlate with the mixing angle.

 For the flavor singlet couplings $\tilde{c}_{d,m}$ we relied on
the relations $\tilde{c}_{d,m} = {c}_{d,m}/\sqrt{3}$ to flavor
octet couplings in the large-$N_c$ limit~\cite{EckerNP321}. One
may argue whether the large-$N_c$ consideration is applicable to
scalar mesons, especially in view of Unitarized ChPT
results~\cite{Pelaez:2003dy}. In this connection, a fit without
any large-$N_c$ restriction would be an extension of the present
approach. However difficulties related to the fitting procedure
may arise, in particular, two more free parameters
$\tilde{c}_{d,m}$ appear, and in view of scarce experimental data
a non-trivial procedure is needed to reduce ambiguities in the
results.

In the present model we obtained the widths of $a_0(980)$ and
$f_0(980)$ decays: $\Gamma_{a_0,tot} = 17.8 \;\text{MeV}$,
$\Gamma_{a_0 \to \pi\eta} = 14.2 \; \text{MeV}$, $\Gamma_{f_0 \to
\pi\pi} = 41.8 \; \text{MeV}$. Many of the calculated observables
are in satisfactory agreement with experiment. At the same time
the calculated ratio ${\Gamma_{\phi \to \gamma f_0}} /
{\Gamma_{\phi \to \gamma a_0}} = 2.64$ only qualitatively agrees
with the experimental value $6.1$. The results of the present
approach are also compared with those of previously developed
kaon-loop model~\cite{Hanhart} and vector-meson-dominance
model~\cite{Black}.

Predictions for the widths of $a_0(980)$ and $f_0(980)$ decays
into $\gamma \rho(770)$ and $\gamma \omega(782)$ are also given
(see Table~\ref{table:width_estim}). The processes, to our
opinion, are of interest for experimental programs in J\"ulich
with COSY~\cite{COSY} and Frascati with DA$\Phi$NE (or its
upgrade)~\cite{RadReturnKLOE,Ambrosino:2006gka}.

Within the present model and one-loop approximation we found
several ratios of the widths which are independent of the
couplings constants. Namely, $\Gamma_{a_0\to\gamma\gamma} /
\Gamma_{\phi\to \gamma a_0} = 0.422$, which is in a qualitative
correspondence with experiment, $\Gamma_{a_0\to\gamma\rho} /
\Gamma_{a_0\to\gamma\omega} =  1.043$ and
$\Gamma_{a_0\to\gamma\rho (\omega)} / \Gamma_{ \phi \to\gamma a_0}
\approx 12$ which have not been tested experimentally yet.

Our calculations show that many predictions are in agreement with
experiment, and therefore support the assumption that $a_0(980)$
and $f_0(980)$ fit in the lightest scalar meson nonet. However, it
is difficult to make an unambiguous conclusion.

The present work makes a solid ground for further studies of
scalar mesons, not only the lightest ones $a_0(980)$ and
$f_0(980)$. The model can be applied in processes of two-photon
production of hadronic states with intermediate scalar resonances.
These processes occur in nucleon-nucleon and electron-positron
collisions (like $e^+e^- \to e^+e^- \pi^+\pi^-$).

\section{Acknowledgement}
We would like to thank N.P.~Merenkov for reading the manuscript
and useful remarks. We are also grateful to S.~Eidelman for
comments and suggestions, to C.~Hanhart and A.~Nefediev for
discussion. The suggestions kindly given by J.R.~Pel\'aez will
help with further development of the approach. One of the authors
(A.Yu.K.) acknowledges support by the INTAS grant 05-1000008-8328
``Higher order effects in $e^+ e^-$ annihilation and muon
anomalous magnetic moment''. S.I. warmly acknowledges the
hospitality of the A.~Salam ICTP in Trieste, Italy, where a part
of the article was prepared.

\appendix{}

\section{Chiral Lagrangian for pseudoscalar and vector mesons}
\label{App:A}
\setcounter{equation}{0}
\def\theequation{A.\arabic{equation}}
\hspace{0.5cm}

In calculations we use $\mathcal{O}(p^{2})$ ChPT Lagrangian for
pseudoscalar mesons $\Phi$, vector mesons and photons, derived by
Ecker~et.~al.~\cite{EckerNP321}, where spin-$1$ mesons are described by
antisymmetric tensor fields $V^{\nu \mu}$. This Lagrangian has
$\mathcal{O}(p^{4})$ chiral power in sense of its equivalence to the ChPT
Lagrangian in which no explicit resonances are introduced
(see~\cite{EckerNP321,EckerPLB223} for details). In the present problem
it is sufficient to keep
\begin{eqnarray}
\label{eq:La}
L^A &=&\frac{f_{\pi }^{2}}{4}\left\langle D_{\mu }UD^{\mu
}U^{\dagger} +\chi U^{\dagger} +\chi^{\dagger} U\right\rangle
-\frac{1}{4}F_{\mu \nu} F^{\mu \nu}
\nonumber \\
&& - \frac{1}{2} \langle \nabla^\lambda V_{\lambda \mu}
\nabla_\nu V^{\nu \mu} - \frac{1}{2}M_V^2 \, V_{\mu \nu} V^{\mu
\nu} \rangle
\nonumber \\
&& +\frac{F_{V}}{2\sqrt{2}}\left\langle V_{\mu \nu }f_{+}^{\mu \nu
}\right\rangle +\frac{ iG_{V}}{\sqrt{2}}\left\langle V_{\mu \nu }u^{\mu }u^{\nu
}\right\rangle
,
\end{eqnarray}
where $\left\langle \cdots \right\rangle$ stands for the trace in flavor
space. The pion weak decay constant  $f_\pi\approx 92.4\;\text{MeV}$,
$F_V$ and $G_V$ are coupling constants. The electromagnetic field $B^\mu$
is included as an external source, $F_{\mu\nu}=\partial_\mu B_\nu -
\partial_\nu B_\mu$ is the electromagnetic field tensor.
The quark mass matrix
 \bgea
 \chi = 2B_0 \; \mathrm{diag} (m_u,m_d,m_s)
 \enea
is expressed in terms of light quark ($q_u, q_d, q_s$) masses and
chiral condensate: \ $ \left\langle0|\bar{q_u}q_u|0\right\rangle =
-f_\pi^2 B_0 \bigl( 1 + \mathcal{O}(m_q) \bigr)$. In the limit of
exact isospin symmetry $\chi = \mathrm{diag}(m_\pi^2,\ m_\pi^2,\ 2
m_K^2-m_\pi^2)$. The terms in Lagrangian relevant for scalar meson
sector are  discussed in Appendix~\ref{App:B}, and are denoted as
$L^B$ in present paper.

Pseudoscalar meson nonet ($J^P=0^-$) contains the $\mathbf{8}_{flavor}$
octet of Goldstone bosons and the $\mathbf{1}_{flavor}$ singlet, namely
$\eta_0$ field. We combine singlet and octet into nonet
following~\cite{Prades}. Thus flavor $SU(3)$ multiplet for pseudoscalar
mesons is
\begin{eqnarray}
\label{eq:appa_def_st}
\Phi &=& \frac{1}{\sqrt{2}} (\pi_1 \lambda_1 + \pi_2 \lambda_2 +
\pi_3 \lambda_3
\nn\\
&&+ K_1 \lambda_4 + K_2 \lambda_5 + K_3 \lambda_6 +
K_4 \lambda_7
\nn\\
&&+ \eta_8 \lambda_8 + \eta_0 \lambda_0),
\end{eqnarray}
where $\lambda_a$ ($a=1,...,8$) are the Gell-Mann matrices, $\lambda_0 =
\sqrt{\frac{2}{3}}\, {1}$, and the physical fields are defined as
 \bgea
\pi^\pm = \frac{1}{\sqrt{2}} (\pi_1 \mp \imath \pi_2), && K^\pm =
\frac{1}{\sqrt{2}} (K_1 \mp \imath K_2),
\\
K^0 = \frac{1}{\sqrt{2}} (K_3 - \imath K_4), && \bar{K^0} =
\frac{1}{\sqrt{2}} (K_3 + \imath K_4), \nn
\\
\pi^0 = \pi_3. \nn
 \enea
Of course such a scheme is well-defined only approximately if the $U(1)$
axial anomaly is neglected. We do not omit problematic $\eta$ meson
within present approach, as it is involved in the dominant decay of $a_0$
meson $a_0 \to \pi \eta$. For $\eta$ - $\eta^\prime$ mixing we choose the
two-parameter scheme~\cite{Beisert}:
 \bgea \label{eq:etamixing}
\eta &=& \cos\theta_8 \,\eta_8 - \sin\theta_0 \,\eta_0 ,\nn\\
\eta^\prime &=& \sin\theta_8 \,\eta_8+ \cos\theta_0 \,\eta_0 .
 \enea
Note that $\eta$ and $\eta^\prime$ are not the orthogonal states.
The angles $\theta_0 = -9.2^\circ$ and $\theta_8 = -21.2^\circ$
are discussed and determined in ~\cite{Beisert,Feldmann} from
experiment.

Further, $f_{+}^{\mu\nu}$ in (\ref{eq:La}) for the external
electromagnetic field reads
 \begin{equation}
f_{+}^{\mu\nu} = e F^{\mu\nu} (u Q u^{+} + u^{+} Q u), \label{eq:f+}
 \end{equation}
where
 \begin{equation}
u \equiv U^{1/2} =  \exp \left( \frac{\ \imath \
\Phi}{\sqrt{2}f_{\pi}} \right)
 \end{equation}
carries non-linear parametrization of the pseudoscalar field. The quark
charge matrix is $Q\equiv
\mathrm{diag}(\frac{2}{3},-\frac{1}{3},-\frac{1}{3})=\frac{1}{2}\lambda_3
+ \frac{1}{2\sqrt{3}}\lambda_8$, and the electron charge is $e =
\sqrt{4\pi\alpha} \approx 0.303$.

The definition of $D_{\mu }U$, $\nabla_\nu V^{\nu \mu}$ and $u^{\nu}$ in
(\ref{eq:La}) can be found in the original work~\cite{EckerNP321}.

From $L^A$~(\ref{eq:La}) the following interactions for physical fields in~$\mathcal{O}(p^2)$
order can be produced~\cite{KorchinIvashyn}:
\begin{eqnarray}
\mathcal{L}_{\gamma P P}& = & -\imath e B_\mu ( \pi^+
\stackrel{\leftrightarrow}{\partial_\mu} \pi^- + K^+
\stackrel{\leftrightarrow}{\partial_\mu} K^- ), \label{eq:F1}
\\
\mathcal{L}_{\gamma \gamma \Phi \Phi} &= & e^2
B^\mu B_\mu (\pi^+\pi^- + K^+K^-),
\label{eq:F2}
\end{eqnarray}
where for any $a$ and $b$ notation $a
\stackrel{\leftrightarrow}{\partial_\mu} b \equiv a\
\partial_\mu \ b - b \ \partial_\mu \ a$
is introduced.
\begin{eqnarray}
\mathcal{L}_{\gamma V} &=&  e F_V  F^{\mu \nu} \bigl(
\frac{1}{2}\rho^0_{\mu\nu} + \frac{1}{6}\omega_{\mu\nu} -
\frac{1}{3\sqrt{2}}\phi_{\mu\nu} \bigr).
\label{eq:F3}
\end{eqnarray}
\begin{eqnarray}
\label{eq:F4}
\mathcal{L}_{VPP}& = & \imath \frac{G_V}{f_\pi^2} \big[ \rho^0_{\mu\nu} ( 2
\partial^\mu\pi^+
\partial^\nu \pi^- + \partial^\mu K^+\partial^\nu K^- )
\nonumber
\\
&&+ \omega_{\mu\nu} \bigl(
\partial^\mu K^+\partial^\nu K^- )
\nonumber
\\
&&+ \phi_{\mu\nu} \bigl( -\sqrt{2}
\partial^\mu K^+\partial^\nu K^- ) \big].
\end{eqnarray}
\begin{eqnarray}
\label{eq:F5} \mathcal{L}_{\gamma V PP} &=& -\frac{e F_V}{f_\pi^2}
\partial^\mu B^\nu \rho_{\mu \nu}^0  \ \pi^+ \pi^-
\\
&& -\frac{e F_V}{2 f_\pi^2}
\partial^\mu B^\nu \left(\rho_{\mu \nu}^0 + \omega_{\mu \nu} - \sqrt{2} \phi_{\mu
\nu}\right)\ K^+ K^- \nn
\\
&&- \frac{2e G_V}{f_\pi^2} B^\nu \rho_{\mu \nu}^0 \left(
\pi^+\partial^\mu \pi^-
 +  \pi^- \partial^\mu\pi^+\right)
\nn
\\
&&- \frac{e G_V}{f_\pi^2} B^\nu \left(\rho_{\mu \nu}^0 +
\omega_{\mu \nu} - \sqrt{2} \phi_{\mu \nu}\right) \nn
\\
&& \times \left( K^+ \partial^\mu K^-  + K^-
\partial^\mu K^+ \right) \nonumber .
\end{eqnarray}

\begin{figure}
\begin{center}
  \includegraphics[width=0.49\textwidth]{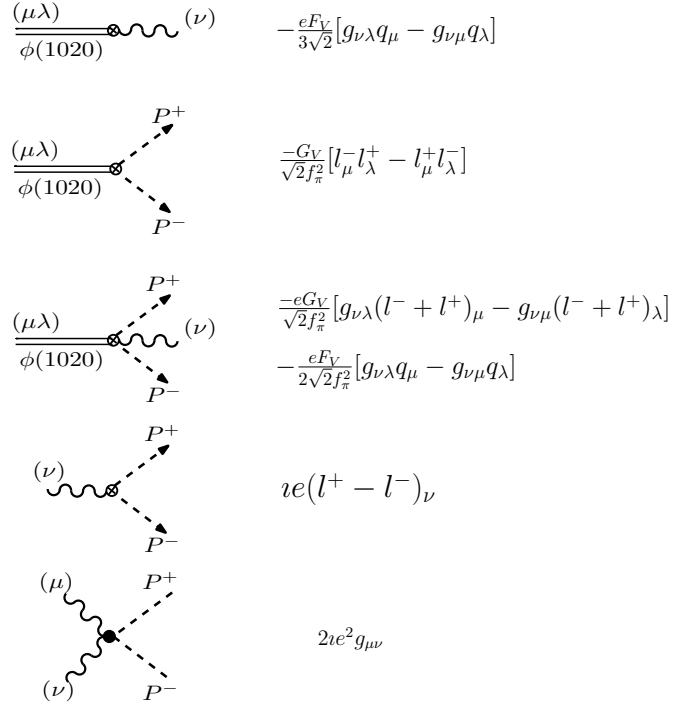}
\end{center}
\caption{The $\mathcal{O}(p^2)$ vertices from  chiral Lagrangian
$L^A$~(\ref{eq:La}). Dash line stands for kaon, double solid - for
vector meson $\phi$, wavy line - for photon. } \label{fig:v2}
\end{figure}

From these terms one can derive the $\mathcal{O}(p^2)$ vertex functions
shown in Fig.~\ref{fig:v2}.

Chiral couplings $F_{V}$ and $G_{V}$ can be extracted from the vector
meson partial widths (see~\cite{EckerNP321,KorchinIvashyn}).
From~(\ref{eq:F3}) and~(\ref{eq:F4}) one calculates the following decay
widths:
 \bgea \label{eq:C1}
 \Gamma_{\rho \to \pi\pi} &=& \frac{G_V^2}{48
\pi f_\pi ^4} \bigl( M_\rho^2 - 4 m_\pi^2 \bigr)^{3/2}
\\
\label{eq:C2} \Gamma_{\rho \to e^+e^-} &=& \frac{e^4 F_V^2}{12 \pi
M_\rho}.
 \enea
For the decays  $\phi \to KK$, $\omega \to e^+e^-$ and $\phi \to
e^+e^-$ one has to take into account the $SU(3)$ relations for
strong and electromagnetic couplings implemented
in~(\ref{eq:F1})-(\ref{eq:F5}). Tables~\ref{table:EM-couplings}
and~\ref{table:strong-couplings} show values for $G_V$ and $F_V$
which are obtained from experimental widths.

\begin{table*}
\caption{Values of electromagnetic coupling constants for vector
mesons}
\label{table:EM-couplings}
\begin{center}
\begin{tabular}{cc|cc|cc}
\hline
\multicolumn{2}{c}{$\rho^0$}  & \multicolumn{2}{c}{$\omega$}    & \multicolumn{2}{c}{$\phi$}              \\
\hline
$\Gamma_{\rho \to e^+ e^-}$, keV & $F_V$~($\text{MeV}$)  & $\Gamma_{\omega \to e^+ e^-}$, keV &  $F_V$~($\text{MeV}$)  & $\Gamma_{\phi \to e^+ e^-}$, keV & $F_V$~($\text{MeV}$)\\
 $7.02 \pm 0.11 $  & $156.162$
& $0.60 \pm 0.02 $  & $137.629$
& $1.27 \pm 0.04 $  & $161.629$ \\
\hline
$\Gamma_{\rho \to \mu^+ \mu^-}$, keV & $F_V$~($\text{MeV}$) & $\Gamma_{\omega \to \mu^+ \mu^-}$, keV & $F_V$~($\text{MeV}$) & $\Gamma_{\phi \to \mu^+ \mu^-}$, keV & $F_V$~($\text{MeV}$)\\
 $6.66 \pm 0.20 $    & $152.358$
& $0.76 \pm 0.26 $      & $154.98$
& $1.21 \pm 0.08 $    & $157.738$\\
\hline
\end{tabular}
\end{center}
\end{table*}

\begin{table}
\caption{Values of vector-meson coupling to two pseudoscalar
mesons (all values are in MeV)} \label{table:strong-couplings}
\begin{center}
\begin{tabular}{r|c|ccc}
\hline
                             &    & $\pi^+ \pi^-$    & $K^+ K^-$         & $K^0 \bar{K}^0$ \\
\hline $\rho^0$
    & exp. width:             & $146.4$& --                               & -- \\
    & $G_V$:             & $65.183         $& --                                 & -- \\
    & $2\ G_V$:          & $130.366        $& --                                 & -- \\
\hline $\omega$
    & exp. width:             & $0.144$ & --                               & -- \\
    &             & (suppressed)&                                & \\
\hline $\phi$
    &  exp. width:            & -- &             $2.096$          & $1.448$ \\
    &  $G_V$:            & -- &             $53.09$                    & $54.45$ \\
    &  $2\ G_V$:         & -- &             $106.18$                 & $108.9$ \\
\hline
\end{tabular}
\end{center}
\end{table}

The condition $F_V = 2\ G_V$ is important for amplitudes to
converge (see Section~\ref{sec:rd2}). From
Tables~\ref{table:EM-couplings} and~\ref{table:strong-couplings}
one sees that this relation is satisfied only approximately, and
the closest values are obtained from (\ref{eq:C1})
and~(\ref{eq:C2}) for the decays $\rho \to \pi^+ \pi^-$ and $\rho
\to e^+ e^-$.


\section{Chiral Lagrangian for light scalar mesons}
\label{App:B}
\setcounter{equation}{0}
\def\theequation{B.\arabic{equation}}
\hspace{0.5cm}

The~$\mathcal{O}(p^2)$ ChPT Lagrangian, which explicitly incorporates
scalar mesons and their interactions with pseudoscalars
reads~\cite{EckerNP321}
 \bgea \label{lagr:general} L^B&=&c_d
\left\langle S^{oct} u_\mu u^\mu \right\rangle + c_m \left\langle S^{oct}
\chi_+ \right\rangle
\nn\\
&&+ \tilde{c}_d S^{sing} \left\langle u_\mu u^\mu \right\rangle
 +  \tilde{c}_m S^{sing} \left\langle  \chi_+ \right\rangle
, \enea $\chi_+ = u^+ \chi u^+ + u\chi u$.
 For other notation and definition see
Appendix~\ref{App:A} and~\cite{EckerNP321}.

Scalar octet $S^{oct}$ and singlet $S^{sing}$ have a priori
independent couplings $c_d, c_m$ and those with hats $\tilde{c}_d,
\tilde{c}_m$. Numerical values of these couplings are determined
by the underlying QCD. However it is difficult to find $c_d$,
$c_m$, $\tilde{c}_d$ and $\tilde{c}_m$ at energies about $1$~GeV
because of the non-perturbative regime of QCD. From assumption of
large number of quark colors ($N_c \to \infty$) it was
shown~\cite{EckerNP321} that octet and siglet (with ``tilde'')
chiral couplings obey relations
 \bgea
 \label{eq:largeNC}
 \tilde{c}_m &=& \mu \frac{c_m}{\sqrt{3}}, \;\;\;\;\;
 \tilde{c}_d = \mu\frac{c_d}{\sqrt{3}}, \;\;\;\;\;
 \mu = \pm 1.
 \enea
Applicability of (\ref{eq:largeNC}) to scalar meson radiative
decays gives rise to some doubts (\cite{Pelaez:2003dy}, for
instance). Anyway we use these constraints to reduce the number of
independent parameters in Section~\ref{sec:generalcalculation}.

For description of scalar meson radiative decays we expand $u_\mu$
in~(\ref{lagr:general}) in series in $\Phi$. The $\mathcal{O}(p^2)$
interaction with scalar mesons is defined by
 \bgea \label{lagr} L^B &=& L^{ChPT}_{oktet} +
L^{ChPT}_{singlet}
,   \\
    L^{ChPT}_{oktet} &=& \frac{2\,c_d}{f_\pi^2} \left\langle S^{oct} \partial_\mu \Phi
    \partial^\mu \Phi  \right\rangle
\nn\\&& -\imath \frac{2e\,c_d}{f_\pi^2}\,B^\mu\,\left\langle
S^{oct} \{\partial_\mu\Phi,[Q,\Phi]\} \right\rangle \nn\\&&
     -\frac{2e^2\,c_d}{f_\pi^2}\,B^\mu B_\mu\, \left\langle S^{oct} [Q,\Phi]^2 \right\rangle
\nn\\
    && - \frac{c_m}{f_\pi^2} \,\left\langle S^{oct} \Phi \{ \chi,\Phi\} \right\rangle
    + \frac{2 c_m}{f_\pi^2} \left\langle S^{oct} \chi \right\rangle
    ,
    \nn\\
L^{ChPT}_{singlet} &=& \frac{2\,\tilde{c}_d}{f_\pi^2} \left\langle \partial_\mu \Phi
\partial^\mu \Phi
\right\rangle   S^{sing} \nn\\&&
 +\imath
\frac{4e\,\tilde{c}_d}{f_\pi^2}\,B^\mu\,\left\langle
\partial_\mu\Phi,[Q,\Phi]
 \right\rangle   S^{sing}
 \nn\\&&
  -\frac{2e^2\,\tilde{c}_d}{f_\pi^2}\,B^\mu B_\mu\, \left\langle  [Q,\Phi]^2 \right\rangle S^{sing}
\nn\\
    && - 2\frac{\tilde{c}_m}{f_\pi^2} \,\left\langle \chi\Phi^2 \right\rangle   S^{sing}
     + \frac{2 \tilde{c}_m}{f_\pi^2} \left\langle \chi \right\rangle S^{sing}
\nn.
 \enea
Apparently Lagrangian (\ref{lagr}) does not yield direct contact coupling
of scalar meson to two photons.

In order to apply (\ref{lagr}) to the physical scalar fields one
has to assume certain multiplet
decomposition~(\ref{eq:multiplet_sc}). Which has to be consistent
with phenomenology. The prominent feature of $a_0$ is its dominant
decay to $\pi\eta$. The KLOE~\cite{Aloiso02C} showed almost no
contribution of $f_0(980)$ resonance in comparison with $a_0(980)$
in the reaction $\phi(1020)\to\gamma\pi^0\eta$. Thus we suppose
that isovector $a_0(980)$ and isoscalar $f_0(980)$ do not mix with
each other. Violation of isospin conservation, related to a
possible $a_0$-$f_0$ mixing, is a subject for a separate work.
This issue can be studied for example by means of the $dd\to$
$({}^4He\ a_0^0 \to)$ ${}^4He\ \pi^0\eta$ reaction at
COSY~\cite{Hanhart_Isomixing}.

As far as we are interested in physical scalar fields, which are
combinations of singlet and octet states, it is convenient to
introduce effective couplings $g_{S\cdots}$ constructed from
constants $c_d, c_m, \tilde{c}_d, \tilde{c}_m$. This allows one to
rewrite Lagrangian in a simpler form.

Let $S$ stand for any scalar field, $a_0$,$f_0$ or $\sigma$, and
$P$ -- for pseudoscalar $\stackrel{\rightarrow}{\pi}= \pi^0,
\pi^\pm$ or $K^\pm$, $K^0$, $\bar{K}^0$. Then
Lagrangian~(\ref{lagr}) can be reduced to
 \bgea
\label{eq:Lb} L^B&=& \frac{1}{f_\pi^2}\sum_{S} S \Bigl[
\frac{g_{S\pi\pi}}{2}\stackrel{\rightarrow}{\pi}^2 +
\frac{g_{S\eta\eta}}{2}\eta^2 + g_{S\pi\eta} \pi^0 \eta
\\&&
+ g_{SKK}(K^+K^- +(-1)^{I_S} K^0\bar{K}^0)
+(\hat{g}_{S\pi\pi}/2)(\partial_\mu\stackrel{\rightarrow}{\pi})^2
\nn\\&& + (\hat{g}_{S\eta\eta}/2)(\partial_\mu\eta)^2 + \hat{g}_{S
\pi^0 \eta}
\partial_\mu\pi^0 \partial^\mu\eta \nn\\&& + \hat{g}_{SKK}(\partial_\mu K^+
 \partial^\mu K^- +(-1)^{I_S} \partial_\mu K^0 \partial^\mu
\bar{K}^0) \nn\\&& + g_{S\gamma\pi\pi}eB_\mu \pi^+
\stackrel{\leftrightarrow}{\partial_\mu} \pi^- + g_{S\gamma
KK}eB_\mu K^+ \stackrel{\leftrightarrow}{\partial_\mu} K^- \nn\\&&
+ g_{S\gamma\gamma\pi\pi}e^2B_\mu B^\mu \pi^+ \pi^- +
g_{S\gamma\gamma KK}e^2B_\mu B^\mu K^+ K^- \Bigr],\nn
 \enea
where $I_S = 0$ for $f_0$, $\sigma$ and $I_S = 1$ for $a_0$. We
introduced the effective couplings $g_{S \pi \pi }$, $g_{S
\eta\eta}$, etc. listed in
Table~\ref{table:generalscalarcouplings}. The couplings which are
absent in Table~\ref{table:generalscalarcouplings} are equal to
zero. In addition, for any scalar meson~$S$ the following
relations for electromagnetic couplings hold
 \bgea
    g_{S\gamma\pi\pi} &=& - \imath \hat{g}_{S\pi\pi} ,\nn\\
    g_{S\gamma KK} &=& - \imath \hat{g}_{SKK} ,\nn\\
    g_{S\gamma\gamma\pi\pi} &=& \hat{g}_{S\pi\pi} ,\nn\\
    g_{S\gamma\gamma KK} &=& \hat{g}_{SKK}
    \label{eq:gghatrelation}.
 \enea

Note also that
 \bgea
    \mathcal{Z}&=&\frac{\cos \theta_0 - \sqrt{2} \sin \theta_8}{\cos( \theta_8 - \theta_0)}
    \nn\\
    &\approx& 1.53,
 \enea
where denominator in $\mathcal{Z}$ is equal to determinant of the
transition matrix~(\ref{eq:etamixing}) from $(\eta_8,\eta_0)$ to
$(\eta,\eta^\prime)$.

\begin{figure}
\begin{center}
\includegraphics[width=0.45\textwidth]{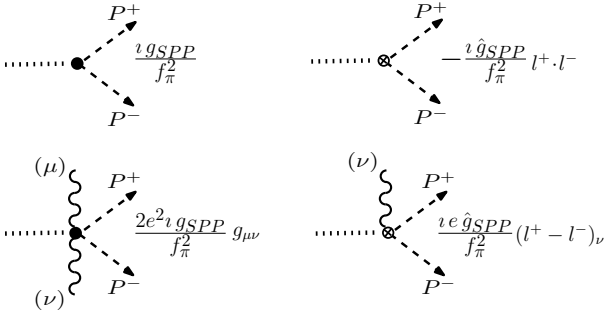}
\end{center}
\caption{The $\mathcal{O}(p^2)$ vertices corresponding to
Lagrangian~(\ref{eq:Lb}). Dotted line stands for scalar meson $S$,
dashed line - for pseudoscalar~$P$. Couplings are shown in
Table~\ref{table:generalscalarcouplings}, see
also~(\ref{eq:gghatrelation}). } \label{fig:v1}
\end{figure}

Lagrangian~(\ref{eq:Lb}) leads to the vertices shown in
Fig.~\ref{fig:v1}.

Table~\ref{table:generalscalarcouplings} shows expressions for the
effective couplings as well as the corresponding chiral powers. Some of
the couplings include the masses of Goldstone bosons and are
$\mathcal{O}(p^2)$, while the others are $\mathcal{O}(p^0)$. Of course
each term in Lagrangian~(\ref{eq:Lb}) carries power $\mathcal{O}(p^2)$.

\begin{table}
\caption{Effective couplings and their chiral powers for scalar mesons}
\label{table:generalscalarcouplings}
\begin{center}
\begin{tabular}{rcl|c}
\hline\noalign{\smallskip}
    $g_{f\pi\pi}$ &=& $- m_\pi^2 (4 \tilde{c}_m \cos \theta - 2\sqrt{2}/\sqrt{3}\, c_m \sin \theta)$, &\\
    $g_{f\eta\eta}$ &=& $- 4/3 \,\tilde{c}_m (4 m_K^2 - m_\pi^2) \cos \theta $&
    \\&&
    $- 2\sqrt{2}/(3\sqrt{3}) \,c_m (8 m_K^2 - 5 m_\pi^2) \sin \theta)$,&$\mathcal{O}(p^2)$\\
    $g_{fKK}$ &=& $- m_K^2(4 \,\tilde{c}_m \cos \theta + \sqrt{2}/\sqrt{3}\, c_m \sin \theta)$ .&\\
\noalign{\smallskip}\hline\noalign{\smallskip}
    $\hat{g}_{f\pi\pi}$ &=& $4 \,\tilde{c}_d \cos \theta - 2\sqrt{2}/\sqrt{3}\, c_d \sin \theta$,& \\
    $\hat{g}_{f\eta\eta}$ &=& $4\, \tilde{c}_d \cos \theta + 2\sqrt{2}/\sqrt{3}\, c_d \sin \theta$ ,&$\mathcal{O}(p^0)$\\
    $\hat{g}_{fKK}$ &=& $4\, \tilde{c}_d \cos \theta + \sqrt{2}/\sqrt{3}\, c_d \sin \theta$.&
\\
\noalign{\smallskip}\hline\noalign{\smallskip}
    $g_{\sigma \pi\pi}$ &=& $- m_\pi^2 (4 \tilde{c}_m \sin \theta + 2\sqrt{2}/\sqrt{3}\, c_m \cos \theta)$,&\\
    $g_{\sigma \eta\eta}$ &=& $- 4/3 \,\tilde{c}_m (4 m_K^2 - m_\pi^2) \sin \theta $&
    \\&&
    $+ 2\sqrt{2}/(3\sqrt{3}) \,c_m (8 m_K^2 - 5 m_\pi^2) \cos \theta)$,&$\mathcal{O}(p^2)$\\
    $g_{\sigma KK}$ &=& $- m_K^2(4 \,\tilde{c}_m \sin \theta - \sqrt{2}/\sqrt{3}\, c_m \cos\theta) $.&\\
\noalign{\smallskip}\hline\noalign{\smallskip}
    $\hat{g}_{\sigma \pi\pi}$ &=& $4 \,\tilde{c}_d \sin \theta + 2\sqrt{2}/\sqrt{3}\, c_d \cos \theta $,&\\
    $\hat{g}_{\sigma \eta\eta}$ &=& $4\, \tilde{c}_d \sin \theta - 2\sqrt{2}/\sqrt{3}\, c_d \cos \theta $,&$\mathcal{O}(p^0)$\\
    $\hat{g}_{\sigma KK}$ &=& $4\, \tilde{c}_d \sin \theta - \sqrt{2}/\sqrt{3}\, c_d \cos \theta $.&
\\
\noalign{\smallskip}\hline\noalign{\smallskip}
    $g_{aKK}$            &=& $- \sqrt{2}\, c_m m_K^2$                                                   ,&$\mathcal{O}(p^2)$\\
    $g_{a\pi\eta}$       &=& $-2 \mathcal{Z}\sqrt{2}/\sqrt{3} \, c_m m_\pi^2$  . &\\
\noalign{\smallskip}\hline\noalign{\smallskip}
    $\hat{g}_{aKK}$      &=& $\sqrt{2} c_d$ ,                                                                        &$\mathcal{O}(p^0)$\\
    $\hat{g}_{a\pi\eta}$ &=& $2 \mathcal{Z} \sqrt{2}/\sqrt{3} \, c_d$ .&
\\
\noalign{\smallskip}\hline
\end{tabular}
\end{center}
\end{table}

Let us now make a remark on relation between the sign of the parameter
$\mu$ in (\ref{eq:largeNC}) and scalar-meson mixing angle $\theta$ in
(\ref{eq:multiplet_sc}).
 As long as present consideration does not involve $\sigma$ meson we can
drop the relation for $\sigma$ in (\ref{eq:multiplet_sc}) and observe a
nontrivial property: the change $\mu \to -\mu$ is equivalent\footnote{
 This equivalence is reflected in
Table~\ref{table:coup_estim_1}.} to the
change $\theta \to \pi - \theta$.

\section{Modifications of the model for virtual photons }
\label{App:C}
\setcounter{equation}{0}
\def\theequation{C.\arabic{equation}}
\hspace{0.5cm}

The complete set of $\mathcal{O}(p^4)$ diagrams has to incorporate
all contributions determined by $\mathcal{O}(p^4)$
Lagrangians~(\ref{eq:F1})-(\ref{eq:F5}) and~(\ref{eq:Lb}).
Lagrangian~(\ref{eq:F3}) generates electromagnetic form factors
(FF's) for pseudoscalar particles inside the loops. These FF's
should replace the tree-level $P P \gamma$ vertices marked by
arrows in Fig.~\ref{fig:p1}, if the FF's do not increase the
chiral power of a diagram and they are calculated from
$\mathcal{O}(p^4)$ Lagrangian.

\begin{figure*}
\begin{center}
\resizebox{0.65\textwidth}{!}{%
  \includegraphics{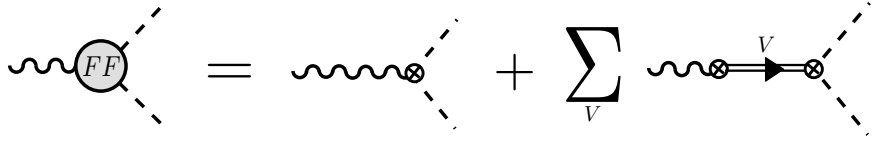}
}
\end{center}
\caption{The $\mathcal{O}(p^2)$ electromagnetic vertex of (off-mass-shell) pseudoscalar meson in
ChPT. All possible intermediate vector resonances $V =\rho^0, \omega, \phi, ...$ in
general contribute.
For real photons only the first term on the r.h.s. is non-zero. }
\label{fig:FFScheme}
\end{figure*}

Note that electromagnetic FF's of kaons and pions have been
studied in various approaches (let us just mention the
considerations for on-mass-shell
pions~\cite{Dubinsky_04,Dubinsky_04_2} and
kaons~\cite{KorchinIvashyn}). The FF calculated from ChPT
Lagrangian includes direct photon--vector meson transition, i.e.
vector meson dominance, as well as ordinary contact interaction
(see illustration in Fig.~\ref{fig:FFScheme} and
Appendix~\ref{App:A}). Fortunately the real photons do not couple
to vector mesons within this approach (see,
e.g.,~\cite{KorchinIvashyn} for discussion of this and one-loop
modification of electromagnetic vertex). Therefore as long as one
is interested in processes with real photons there are no
$\mathcal{O}(p^4)$ diagrams additional to those shown in
Fig.~\ref{fig:p1}, and therefore Fig.~\ref{fig:p1} gives the
complete set of diagrams in this order. The similar reasoning
applies to consideration of diagrams shown in Fig.~\ref{fig:p2}
and Fig.~\ref{pic:p3} for $\phi \to \gamma a_0/f_0$ and $f_0/a_0
\to \gamma\ \rho/\omega$ decay respectively.

\section{Dimensional regularization. Loop integrals}
\label{App:D}
\setcounter{equation}{0}
\def\theequation{D.\arabic{equation}}
\hspace{0.5cm}

In calculation of loop integrals we use the dimensional regularization
method (see, for instance,~\S~7 and Appendix~\ref{App:B} in~\cite{DR}).

The dimension of space-time $D=4-2\epsilon$ in the limit $\epsilon\to 0$
corresponds to that of 4-dimensional Minkowsky space. In the text this
limit is assumed in all expressions.
Integration measure for 4-dimensional space is replaced by that for
$D$-dimensional space: $d^4q \to (\Lambda^2)^\epsilon d^Dq$, where
arbitrary regularization parameter $\Lambda$ has units of mass. Integrals
with this measure are defined via the analytical continuation from the
space with the integer number of dimensions. Metric tensor obeys the
condition $g^{\mu\nu}g_{\mu\nu}= D$.
The Dirac matrices satisfy the anti-commutation relations
$\{\gamma^\mu,\gamma^\nu\} = 2 g^{\mu\nu}$. Here $\gamma^\mu \gamma_\mu =
D$, and the ordinary trace formulae are generalized to
$\mathrm{Tr}(\gamma^\mu\gamma^\nu)=2^{D/2} g^{\mu\nu} $,~etc.

\begin{table}
\caption{Table of typical $D$-dimensional integrals (for any vector
$Q^\mu$ and complex number $R$).} \label{table:regulariz}
\begin{center}
\begin{tabular}{rl}
\hline\noalign{\smallskip} $ \Lambda^{2\epsilon} \int \frac{
d^Dl}{(2\pi)^D} \frac{1}{l^2 + R} =$& $ \frac{-\imath R}{(4\pi)^2}
[I_\epsilon + 1 - \ln (\frac{-R}{\Lambda^2} ) ] $
,\\
$ \Lambda^{2\epsilon} \int \frac{ d^Dl}{(2\pi)^D} \frac{\{1, l^\mu l^\nu
\} }{(l^2-2l\!\cdot\!Q + R)^2} =$ & $ \{1, Q^\mu Q^\nu \}
\frac{\imath}{(4\pi)^2}$
\\
&
$\times\ [I_\epsilon - \ln (\frac{Q^2-R}{\Lambda^2} ) ] $
\\
& $+ \; \{ 0, g ^{\mu\nu} \} \frac{\imath}{32\pi^2}(Q^2 - R)$
\\
&
$\times\ [I_\epsilon + 1 - \ln (\frac{Q^2-R}{\Lambda^2} ) ] $
,\\
$ \Lambda^{2\epsilon}  \int \frac{ d^Dl}{(2\pi)^D} \frac{\{1,l^\alpha,
l^\alpha l^\beta, l^\alpha l^\beta l^\nu \}}{(l^2-2l\!\cdot\!Q + R)^3}
=$& $\frac{-\imath}{2(4\pi)^2} \frac{\{ 1, Q^\alpha, Q^\alpha Q^\beta,
Q^\alpha Q^\beta Q^\nu \}}{Q^2 - R}$
\\
&
 $+ \frac{\imath}{4(4\pi)^2}[I_\epsilon -
 \ln (\frac{Q^2-R}{\Lambda^2} )]$
\\ &
$\times \{ 0,0,g^{\alpha\beta}, (g^{\alpha\beta} Q^\nu $
 \\
&
$+ g^{\alpha\nu} Q^\beta + g^{\nu\beta} Q^\alpha )\}$
,
\\
\noalign{\smallskip}\hline
\end{tabular}
\end{center}
\end{table}

In calculation of loop diagrams the typical integrals presented in
Table~\ref{table:regulariz} arise. For divergent terms we define
 \bgea
 I_\epsilon &=& 1/\epsilon - \gamma_\epsilon + \ln 4\pi .\nn
 \enea
Euler-Mascheroni constant $\gamma_\epsilon \approx 0.57721566490$ can be expressed
in terms of gamma-function derivative $\gamma_\epsilon = -
\Gamma^\prime(1) = -\int_0^\infty dx \exp{(-x)} \ln x $.

%
%

\end{document}